\documentclass[11pt,twoside]{article}

\usepackage{fancyhdr}
\usepackage{epsfig}
\usepackage{latexsym}
\usepackage{microtype}
\usepackage{lastpage} 
\usepackage{hyperref}
\usepackage[utf8]{inputenc}
\usepackage{%afterpage,
enumitem,
mathpartir,%epigraph,
amsmath,amsfonts,%amsthm,
amssymb,%
stmaryrd,%cmll,% microtype,
graphicx,empheq,wrapfig,
yhmath,mathabx
} %
\usepackage[mathcal]{euscript}
\usepackage{common/fullshort}
\setboolean{fullpaper}{true}
\newlist{myaxioms}{enumerate}{10}
\setlist[myaxioms,1]{label=(P\arabic*)}
\setlist[myaxioms]{resume}

\newlist{enumeratei}{enumerate}{10}
\setlist[enumeratei]{label=\emph{(\roman*)}}

\DeclareMathOperator{\ob}{ob}
\DeclareMathOperator{\mor}{mor}

\DeclareMathOperator{\dom}{dom}
\DeclareMathOperator{\cod}{cod}

\DeclareMathOperator{\colim}{colim}

\DeclareMathOperator{\Fam}{Fam}

\newcommand{\restr}[2]{#1_{|#2}}

\newcommand{\bigcat}[1]{\ensuremath{\mathsf{#1}}}
\newcommand{\maji}[1]{\ensuremath{\mathbb{#1}}}

\newcommand{\Vars}{\bigcat{Vars}}
\newcommand{\Names}{\bigcat{Names}}

\newcommand{\commentthis}[1]{}

\newcommand{\Cospan}[1]{\bigcat{Cospan} (#1)}

\newcommand{\into}{\hookrightarrow}
\newcommand{\otni}{\hookleftarrow}

\newcommand{\ot}{\leftarrow}
\newcommand{\xto}[1]{\xrightarrow{#1}}
\newcommand{\xot}[1]{\xleftarrow{#1}}

\newcommand{\xinto}[1]{\xhookrightarrow{#1}}

\newcommand{\eg}{e.g.,}

\renewcommand{\hat}[1]{\widehat{#1}}
\DeclareMathOperator{\Ran}{Ran}

\def\framed{%
\setbox0=\vbox\bgroup%
\advance\hsize by -2\fboxsep\advance\hsize by -2\fboxrule%
\linewidth=\hsize%
}
\def\endframed{%
\egroup\noindent\framebox[\textwidth]{\box0}\vspace*{1mm}}

\usepackage{tikz,common/tikz-cats}
\tikzset{todim/.style = {decoration={markings, mark=at position .5 with %
      {\draw (-1pt,-1pt) rectangle (1pt,1pt);}},postaction={decorate}}}
\renewcommand{\diaggrandhauteur}{.6}
\renewcommand{\diaggrandlargeur}{1.3}

\usepackage{color}
\definecolor{dkgreen}{rgb}{0,0.2,0}

\newcommand{\B}{\maji{B}}

\newcommand{\C}{\maji{C}}

\newcommand{\tick}{\daimon}

\newcommand{\Chat}{\hat{\maji{C}}}

\newcommand{\CE}{\E}

\newcommand{\CV}{\V}

\newcommand{\CVX}{\CV_X}
\newcommand{\CVY}{\CV_Y}
\newcommand{\CVn}{\CV_{[n]}}

\newcommand{\CVni}{\CV_{[n']}}

\newcommand{\CEXhat}{\widehat{\CE_X}}

\newcommand{\CVnhat}{\widehat{\CV_{[n]}}}

\newcommand{\CVXhat}{\widehat{\CVX}}

\newcommand{\CVYhat}{\widehat{\CVY}}
\newcommand{\CVnhatf}{\OPsh{\CVn}}

\newcommand{\CVhatf}{\OPsh{\CV_{[-]}}}

\newcommand{\CEX}{\CE_X}

\newcommand{\D}{\maji{D}}

\newcommand{\E}{\maji{E}}

\newcommand{\Eh}{\B}

\newcommand{\F}{\maji{F}}

\newcommand{\G}{\maji{G}}
\newcommand{\GG}{\bigcat{G}}

\newcommand{\K}{\maji{K}}

\newcommand{\MMM}{\mathcal{M}}

\newcommand{\V}{\maji{V}}

\newcommand{\W}{\maji{W}}

\newcommand{\CW}{\W}

\newcommand{\CWofX}{\CW(X)}

\newcommand{\CWofXhat}{\widehat{\CWofX}}

\renewcommand{\SS}{\bigcat{S}}

\newcommand{\SSn}{\SS_{[n]}}

\newcommand{\SSX}{\SS_X}

\renewcommand{\tt}{\bigcat{t}} %

\newcommand{\Set}{\bigcat{Set}}

\newcommand{\FinOrd}{\bigcat{FinOrd}}
\newcommand{\OPsh}[1]{\wideparen{#1}}

\newcommand{\Cat}{\bigcat{Cat}}

\newcommand{\CAT}{\bigcat{CAT}}

\newcommand{\trou}{\boxempty}

\newcommand{\Nat}{\mathbb{N}} % set of natural numbers

\newcommand{\lra}{\textsc{lra}}

\newcommand{\un}{I}
\newcommand{\para}{\mathbin{\mid}}

\newcommand{\transl}[1]{\llbracket #1 \rrbracket}

\newcommand{\rond}{\circ}
\newcommand{\vrond}{\mathbin{\scriptstyle \bullet}}

\newcommand{\id}{\mathit{id}}
\newcommand{\iso}{\cong}

\newcommand{\impll}{\multimap}
\newcommand{\tens}{\otimes}

\newcommand{\card}[1]{|#1|}

\newcommand{\eq}{\mathrel{\texttt{:=}}}
\newcommand{\ens}[1]{\{ #1 \}}
\newcommand{\inv}[1]{#1^{-1}}
\newcommand{\aalt}{\mathrel{|}}
\newcommand{\name}[1]{\ulcorner #1 \urcorner}
\newcommand{\op}[1]{#1^{\mathit{op}}}

\newcommand{\subs}[1]{[#1]}

\newcommand{\Gam}{\Gamma}
\newcommand{\Del}{\Delta}

\newcommand{\equi}{\simeq}

\newcommand{\bureaucratic}[1]{}

\newcommand{\abar}{\overline{a}}
\newcommand{\bbar}{\overline{b}}

\newcommand{\sender}{\epsilon}
\newcommand{\receiver}{\rho}

\newcommand{\iotaof}[2]{\iota_{#1,#2}}

\newcommand{\iotapos}[1]{o_{#1}}
\newcommand{\iotaneg}[1]{\iota_{#1}}
\newcommand{\iotaposni}{\iotapos{n,i}}

\newcommand{\iotaposnj}{\iotapos{n,j}}
\newcommand{\iotanegni}{\iotaneg{n,i}}
\newcommand{\iotanegmj}{\iotaneg{m,j}}
\newcommand{\iotanegnj}{\iotaneg{n,j}}
\newcommand{\paraof}[1]{\pi_{#1}}

\newcommand{\paralof}[1]{\pi^l_{#1}}
\newcommand{\pararof}[1]{\pi^r_{#1}}
\newcommand{\paran}{\paraof{n}}
\newcommand{\paraln}{\paralof{n}}
\newcommand{\pararn}{\pararof{n}}

\newcommand{\paraofij}[2]{\mathbin{{}_{#1}\!\para\!{}_{#2}}}
\newcommand{\paraij}{\paraofij{i}{j}}

\newcommand{\paradeuxun}{\paraofij{2}{1}}

\newcommand{\nuof}[1]{\nu_{#1}}
\newcommand{\nun}{\nuof{n}}

\newcommand{\tickof}[1]{\tick_{#1}}
\newcommand{\tickn}{\tickof{n}}

\newcommand{\tauof}[4]{\tau_{#1,#2,#3,#4}}
\newcommand{\taunimj}{\tauof{n}{i}{m}{j}}

\newcommand{\SynF}{\bigcat{F}}
\newcommand{\derivmap}{\bigcat{d}}

\newcommand{\Pl}{\mathrm{Pl}}

\newcommand\independent{\protect\mathpalette{\protect\independenT}{\perp}} 
\def\independenT#1#2{\mathrel{\rlap{$#1#2$}\mkern2mu{#1#2}}} 
\newcommand{\bbot}{\mathord{\independent}}

\newcommand{\daimon}{\heartsuit}

\newcommand{\recin}[2]{\texttt{rec}\ #1\ \texttt{in}\ #2}
\newcommand{\Gl}{\GG}
\newcommand{\GlX}{\Gl_X}

\newcommand{\SStoGG}{\bigcat{Gl}}

\newcommand{\state}{\sigma}
\newcommand{\statei}{\sigma'}

\newcommand{\floor}[1]{\lfloor #1 \rfloor}
\newcommand{\deriv}{\partial}

\newcommand{\extafun}[1]{\overline{(-)}}

\newcommand{\Exonze}{\mathrm{Ex}_{11}}

\tikzset{history/.style = {-open triangle 45}}

\newenvironment{myproof}[1][Proof]{\par\noindent {\bf #1:}\ \ \ }{ \hfill$\Box$\hspace{1ex}}
\usepackage{common/sacs-preamble}

\newcommand{\proof}[1]{\par\noindent {\bf Proof:}\ \ \ #1 \hfill$\Box$\hspace{1ex}}

\newcommand{\TheAuthor}{}
\newcommand{\Author}[1]{\renewcommand{\TheAuthor}{#1}}

\newcommand{\TheTitle}{}
\newcommand{\Title}[1]{\renewcommand{\TheTitle}{#1}}

\Author{Tom Hirschowitz and Damien Pous}
\Title{Innocent strategies as presheaves \\ and interactive equivalences for CCS}
\begin{document}

\parindent=8mm
\noindent

% \underline{{\noindent \bf \scriptsize Scientific Annals of Computer Science vol. 21, 2011, pp. \thepage--\hyperlink{lastpage}{\pageref{LastPage}}}}
% \vskip -3mm
% \noindent
% \vskip -1mm
% \noindent
% 

\vspace{1cm}
\begin{center}
{\Large\bf \TheTitle}\footnote{Both authors have been partially funded by the French projects CHoCo
(ANR-07-BLAN-0324), PiCoq (ANR-10-BLAN-0305-01), and CNRS PEPS CoGIP.}
\end{center}
\vspace{4mm}

\begin{center}
{\large Tom HIRSCHOWITZ}\footnote{CNRS, Universit\'e de Savoie, France, \texttt{tom.hirschowitz@univ-savoie.fr}}
and 
{\large Damien POUS}\footnote{CNRS, Laboratoire d'Informatique de Grenoble, France, \texttt{damien.pous@ens-lyon.fr}}
\end{center}
\vspace{3ex}

\date{}

\begin{abstract}                                          
  Seeking a general framework for reasoning about and comparing
  programming languages, we derive a new view of Milner's
  CCS~\cite{Milner80}.  We construct a category $\CE$ of \emph{plays},
  and a subcategory $\CV$ of \emph{views}.  We argue that presheaves
  on $\CV$ adequately represent \emph{innocent} strategies, in the
  sense of game semantics~\cite{hyland97}. We equip innocent
  strategies with a simple notion of interaction.

  We then prove decomposition results for innocent strategies, and,
  restricting to presheaves of finite ordinals, prove that innocent
  strategies are a final coalgebra for a polynomial
  functor~\cite{Kock01012011} derived from the game. This leads to a
  translation of CCS with recursive equations.

  Finally, we propose a notion of \emph{interactive equivalence} for
  innocent strategies, which is close in spirit to Beffara's
  interpretation~\cite{bef05:lrc} of testing
  equivalences~\cite{DBLP:journals/tcs/NicolaH84} in concurrency
  theory. In this framework, we consider analogues of \emph{fair}
  testing and \emph{must} testing. We show that must testing is
  strictly finer in our model than in CCS, since it avoids what we
  call `spatial unfairness'. Still, it differs from fair testing, and
  we show that it coincides with a relaxed form of fair testing.
\end{abstract}

\newcommand{\diaghautici}{.05} \newcommand{\diaglargeici}{.5} %

\paragraph{Note:} This is an expanded version of our ICE '11
paper~\cite{HP11}. It notably simplifies a few aspects of the
development, and corrects the mistaken statement that fair and must
testing coincide in our semantic framework. Must testing only
coincides with a relaxed variant of fair testing. This version also
subsumes a previous preprint, providing more compact proofs.

\section{Overview}

\paragraph{Theories of programming languages}
Research in programming languages is mainly technological. Indeed, it
heavily relies on techniques which are ubiquitous in the field, but
almost never formally made systematic. Typically, the definition of a
language then quotiented by variable renaming ($\alpha$-conversion)
appears in many theoretical papers about functional programming
languages. Why isn't there yet any abstract framework performing these
systematic steps for you?  Because the quest for a real \emph{theory}
of programming languages is not achieved yet, in the sense of a corpus
of results that actually help developing them or reasoning about them.
However, many attempts at such a theory do exist.

A problem for most of them is that they do not account for the
dynamics of execution, which limits their range of application. This
is for example the case of Fiore et al.'s second-order
theories~\cite{DBLP:conf/lics/Fiore08,hirscho:lam,DBLP:journals/iandc/HirschowitzM10}.
A problem for most of the other theories of programming languages is
that they neglect denotational semantics, i.e., they do not provide a
notion of model for a given language.  This is for example the case of
Milner et al.'s \emph{bigraphs}~\cite{Milner:bigraphs}, or of most
approaches to \emph{structural operational
  semantics}~\cite{PlotkinSOS}, with the notable exception of the
\emph{bialgebraic semantics} of Turi and
Plotkin~\cite{plotkin:turi:bialgebraic}.  A recent, related, and
promising approach is \emph{Kleene coalgebra}, as advocated by
Bonsangue et al.~\cite{DBLP:conf/fossacs/BonsangueRS09}.  Finally,
\emph{higher-order rewriting}~\cite{DBLP:conf/lics/Nipkow91}, and its
semantics in double categories~\cite{DBLP:conf/birthday/GadducciM00}
or in cartesian closed
2-categories~\cite{HIRSCHOWITZ:2010:HAL-00540205:2}, is not currently
known to adequately account for process calculi.

\paragraph{Towards a new approach}
The most relevant approaches to us are bialgebraic semantics and
Kleene coalgebra, since the programme underlying the present paper
concerns a possible alternative.  A first difference, which is a bit
technical but may be of importance, is that both bialgebraic semantics
and Kleene coalgebra are based on labelled transition systems (LTSs),
while our approach is based on reduction semantics. This seems
relevant, since reduction semantics is often considered more primitive
than LTSs, and much work has been devoted to deriving the latter from
the
former~\cite{DBLP:conf/concur/Sewell98,DBLP:conf/concur/LeiferM00,Milner:bigraphs,Sobocinski:grpos,modularLTS}.

More generally, our approach puts more emphasis on interaction between
programs, and hence is less interesting in cases where there is no
interaction. A sort of wild hope is that this might lead to unexpected
models of programming languages, e.g., physical ones. This could also
involve finding a good notion of morphism between languages, and
possibly propose a notion of compilation.  At any rate, the framework
is not set up yet, so investigating the precise relationship with
bialgebraic semantics and Kleene coalgebra is deferred to further
work.

How will this new approach look like? Compared to such long-term
goals, we only take a small step forward here, by considering a
particular case, namely Milner's CCS~\cite{Milner80}, and providing a
new view of it.  This view borrows ideas from the following lines of
research: game semantics~\cite{hyland97}, and in particular the notion
of an \emph{innocent strategy}, \emph{graphical
  games}~\cite{Miller08,DBLP:journals/entcs/HirschowitzHH09}, Krivine
realisability~\cite{DBLP:journals/tcs/Krivine03},
ludics~\cite{DBLP:journals/mscs/Girard01}, testing equivalences in
concurrency~\cite{DBLP:journals/tcs/NicolaH84,bef05:lrc}, the presheaf
approach to
concurrency~\cite{DBLP:conf/lics/JoyalNW93,DBLP:journals/mscs/KasangianL99},
and sheaves~\cite{MM}. It is also, more remotely, related to graph
rewriting~\cite{Ehrig} and
computads~\cite{DBLP:journals/tcs/Burroni93}.

\paragraph{From strategies to presheaves} Game
semantics~\cite{hyland97} has provided fully complete models of
programming languages.  It is based on the notion of a
\emph{strategy}, i.e., a set of \emph{plays} in some game, satisfying a
few conditions. In concurrency theory, taking as a semantics the set
of accepted plays, or `traces', is known as \emph{trace
  semantics}. Trace semantics is generally considered too coarse,
since it equates, for a most famous example, the right and the wrong
coffee machines, $a.(b+c)$ and $ab + ac$~\cite{Milner80}.

An observation essentially due to Joyal, Nielsen, and Winskel is that
strategies, i.e., prefix-closed sets of plays, are actually particular
\emph{presheaves of booleans} on the category $\C$ with plays as
objects, and prefix inclusions as morphisms. By presheaves of booleans
on $\C$ we here mean functors $\op\C \to 2$, where $2$ is the preorder
category $0 \leq 1$. If a play $p$ is \emph{accepted}, i.e., mapped to
$1$, then its prefix inclusions $q \into p$ are mapped to the unique
morphism with domain $1$, i.e., $\id_1$, which entails that $q$ is
also accepted.

Following Joyal, Nielsen, and Winskel, we observe that considering
instead presheaves (of sets) on $\C$ yields a much finer
semantics. So, a play $p$ is now mapped to a set $S(p)$, to be thought
of as the set of ways for $p$ to be accepted by the strategy
$S$. Considering the set of players as a team, $S(p)$ may also be
thought of as the set of possible \emph{states} of the team after
playing $p$ -- which is empty if the team never accepts to play $p$.

This presheaf semantics is fine enough to account for
bisimilarity~\cite{DBLP:conf/lics/JoyalNW93,DBLP:journals/mscs/KasangianL99}.
Indeed, presheaves are essentially forests with edges labelled by
moves. For example, in the setting where plays are finite words on an
alphabet, the wrong coffee machine may be represented by the presheaf
$S$ defined by the equations on the left and pictured as on the right:
\begin{center}
  \begin{center}
    \begin{minipage}[c]{0.23\linewidth}       
        $\begin{array}[t]{l}
       S (\epsilon) = \ens{\star}, \\
       S (a) = \ens{x,x'}, \\
       S (ab) = \ens{y}, \\
       S (ac) = \ens{y'},
     \end{array}$
   \end{minipage} \hfil \begin{minipage}[c]{0.42\linewidth}
     $\begin{array}[t]{l}
       S (\epsilon \into a) = \ens{x \mapsto \star, x' \mapsto \star}, \\
       S (a \into ab) = \ens{y \mapsto x}, \\
       S (a \into ac) = \ens{y' \mapsto x'}:
    \end{array}$
    \end{minipage}
      \hfil
    \diag(.15,.5){%
      \& |(root)| \star \& \\
      |(x)| x \& \& |(x')| x' \\
      \\
      |(y)| y \& \& |(y')| y'. %
    }{%
      (root) edge[-,labelal={a}] (x) %
      edge[-,labelar={a}] (x') %
      (x)  edge[-,labell={b}] (y) %
      (x')  edge[-,labelr={c}] (y') %
    }
  \end{center}
\end{center}
So, in summary: the standard notion of strategy may be generalised to
account for branching equivalences, by passing from presheaves of
booleans to presheaves of sets.

\paragraph{Multiple players}
Traditional game semantics mostly emphasises two-player
games. There is an implicit appearance of three-player games in the
definition of composition of strategies, and of four-player games in
the proof of its associativity, but these games are never given a
proper status.  A central idea of graphical games, and to a lesser
extent of ludics, is the emphasis on multiple-player games.

Here, there first is a base category $\B$ of \emph{positions}, whose
objects represent configurations of players.  Since the game
represents CCS, it should be natural that players are related to each
other via the knowledge of \emph{communication channels}.  So,
roughly, positions are bipartite graphs with vertex sets
\emph{players} and \emph{channels}, and edges from channels to players
indicating when the former is known to the latter.  As a first
approximation, morphisms of positions may be thought of as just
embeddings of such graphs.

Second, there is a category $\E$ of \emph{plays}, with a functor to
$\B$ sending each play to its initial position. Plays are represented
in a more flexible way than just sequences of moves, namely using a
kind of string diagrams. This echoes the idea~\cite{Mellies04} that
two moves may be independent, and that plays should not depend on the
order in which two independent moves are performed. Furthermore, our
plays are a rather general notion, allowing, e.g., to focus on a given
player.
% to look at how only some players of the initial position evolve.
% \tom{j'essaie d'expliquer pour qu'on reformule:
%   je voulais dire que notre notion de partie est plus générale que
%   juste suivre l'évolution d'un groupe de joueurs; elle permet
%   d'oublier de regarder certains joueurs, voire un morceau de coup, cf
%   l'inclusion du fork gauche dans le fork plein. Oui?}
%\dam{j'ai reformulé, mais sans mettre l'idée du half-fork : c'est trop
% tôt}
%
 Morphisms of plays account both for:
\begin{itemize}
\item prefix inclusion, i.e., inclusion of a play into a longer play, and
\item position enlargement, e.g., inclusion of information about some
  players into information about more players.
\end{itemize}

Now, restricting to plays over a given initial position $X$, and then taking
presheaves on this category $\E_X$, we have a category of strategies
on $X$.

\paragraph{Innocence}
A fundamental idea of game semantics is the notion of
\emph{innocence}, which says that players have a restricted
\emph{view} of the play, and that their actions may only depend on
that view. 

We implement this here by defining a subcategory $\CVX \into \CEX$ of
\emph{views} on $X$, and deeming a presheaf $F$ on $\E_X$
\emph{innocent} when it is determined by its restriction $F'$ to
$\CVX$, in the sense that it is isomorphic to the \emph{right Kan
  extension}~\cite{MacLane:cwm} of $F'$ along $\op{\CVX} \into \op{\CEX}$.

We then define \emph{innocent strategies} to be just presheaves on
$\CVX$, and view them as (naive) strategies via the (essential)
embedding $\CVXhat \into \CEXhat$ induced by right Kan extension.

\paragraph{Interaction}
For each position $X$, we thus have a category $\SS_X = \CVXhat$ of
innocent strategies.  In game semantics, composition of strategies is
achieved in two steps: \emph{interaction} and \emph{hiding}.
Essentially, interaction amounts to considering the three-player game
obtained by letting two two-player games interact at a common
interface. Hiding then forgets what happens at that interface, to
recover a proper two-player game.

We have not yet investigated hiding in our approach, but, thanks to
the central status of multiple-player games, interaction is accounted
for in a very streamlined way.  For any position $X$ with two
subpositions $X_1 \into X$ and $X_2 \into X$ such that each player is
in either $X_1$ or $X_2$, but none is in both, given innocent
strategies $F_1 \in \SS_{X_1}$ and $F_2 \in \SS_{X_2}$, there is a
unique innocent strategy, the \emph{amalgamation} $[F_1, F_2]$ of
$F_1$ and $F_2$, whose restrictions to $X_1$ and $X_2$ are $F_1$ and
$F_2$.

Amalgamation in this sense models interaction in the sense of game
semantics, and, using the correspondence with presheaves on $\CEX$
given by right Kan extension, it is the key to defining interactive
equivalences.

\paragraph{CCS}
Next, we define a translation of CCS terms with recursive equations
into innocent strategies. This rests on \emph{spatial} and
\emph{temporal} decomposition results for innocent strategies. Spatial
decomposition says that giving a strategy on a position $X$ is the
same as giving a strategy for each of its players. Temporal
decomposition says that a strategy is determined up to isomorphism by
its set of initial states, plus what remains of each of them after
each basic move. Restricting to presheaves of finite ordinals, we also
prove that innocent strategies form a final coalgebra for a polynomial
functor (in the sense of Kock~\cite{Kock01012011}) derived from the
game, thus hinting at links with Kleene coalgebra. It is then easy to
translate finite CCS into the language induced by our polynomial
functor, and to finally extend the translation to CCS with recursive
equations via infinite unfolding.

A natural question is then: which equivalence does this translation
induce on CCS terms? As explained in the following paragraph, we
provide some preliminary results about interactive equivalences, but
essentially leave the question open.

% \tom{si le
%   temps le permet, ajouter ce qu'on sait dire de la traduction. pour
%   l'instant pas grand-chose, mais je crois savoir démontrer que la
%   congruence structurelle est préservée (au sens des tests). a mon
%   avis, vu comme c'est parti, on n'aura pas le temps.}  \dam{oui: pas
%   gagné, et pas sûr que ça soit très sexy de dire ``trop balaises, la
%   congruence structurelle est préservée'' (même si ce n'est pas
%   immédiat...)}

\paragraph{Interactive equivalences}
Returning to our model, we then define a notion of \emph{interactive
  equivalence}, which is close in spirit to both testing equivalences
in concurrency theory and Krivine realisability and ludics.

The game, as sketched above, allows interacting with players which are
not part of the considered position. E.g., a player in the considered
position $X$ may perform an input which is not part of any
synchronisation. A \emph{test} for an innocent strategy $F$ on $X$ is
then, roughly, an innocent strategy $G$ on a position $X'$ with the
same channels as $X$.  To decide whether $F$ \emph{passes} the test
$G$, we consider a restricted variant of the game on the `union' $X
\cup X'$, forbidding any interaction with the outside. We call that
variant the \emph{closed-world} game.
% \dam{question générale : pourquoi le
%   ``closed-world'' n'est pas un cas particulier de hiding} \tom{le
%   hiding prend une partie sur une position avec plusieurs joueurs, qui
%   peuvent interagir avec l'extérieur et renvoie une partie sur une
%   position avec un seul joueur, que l'extérieur ne peut pas distinguer
%   de la partie originale; le closed-world restreint la notion de
%   partie en interdisant toute interaction avec l'extérieur.}

Then $F$ passes $G$ iff the amalgamation $[F, G]$, right Kan extended
to $\CE_{X \cup X'}$ and then restricted to the closed-world game,
belongs to some initially fixed class of strategies, $\bbot_{X \cup
  X'}$. Finally, two innocent strategies $F$ and $F'$ on $X$ are
equivalent when they pass the same tests.

Here are two examples for $\bbot$. Consider a \emph{tick} move,
fixed in advance. Then call \emph{successful} all plays containing at
least one tick, and accordingly call successful all states reached
after a successful play. One may consider:
\begin{itemize}
\item $\bbot^m$, consisting of strategies whose maximal states (those
  that admit no strict extensions) are all successful; the tick move
  plays a r\^ole analogous to the daimon in ludics: it is the only
  move which is observable from the outside;
\item $\bbot^f$, consisting of strategies in which all states
  on \emph{finite} plays admit a successful extension.
% \item $\bbot^e$, consisting of strategies in which all states
%   admit a successful extension.
\end{itemize}
From the classical concurrency theory point of view on behavioural
equivalences, the first choice clearly mimicks \emph{must} testing
equivalence, while the second mimicks \emph{fair} testing
equivalence~\cite{DBLP:conf/icalp/NatarajanC95,DBLP:conf/concur/BrinksmaRV95}.
% \dam{avec le bib que j'ai (../common/bib.bib), ces deux refs sont manquantes}
% The third choice is
% irrelevent in standard trace models since an infinite trace cannot be
% extended: $\bbot^e$ and $\bbot^m$ are equivalent when considering
% traces, the restriction to finite traces the $\bbot^e$ is crucial to
% get fair testing.

Consider the processes $\Omega$ and $\Omega|\abar$, where $\Omega$ is
a process doing infinitely many silent transitions.  These processes
are intuitively quite different: the latter can do an output on the
channel $a$, while the former cannot. They are however equated by
standard must testing equivalence: the infinite trace provided by
$\Omega$ may prevent the output prefix from being performed. In fact,
must testing equivalence heavily relies on the potential unfairness of
the scheduler. In the literature, this peculiar behaviour actually
motivates the introduction of fair testing equivalence.

% \tom{est-ce qu'on
%   pourrait pas présenter tout ça en partant de (1) dans le cas
%   standard, on a f, m et sf; (2) sf coïncide trivialement avec m; (3)
%   dans notre cadre on peut imiter f, m et sf, mais c'est plus du tout
%   évident de les relier; (4) m est moins con, exemple à l'appui, donc
%   on espère m = f; (5) mais on montre m = sf, encore, et m $\neq$ f,
%   encore.}
% \dam{le problème avec ça, c'est que sf n'est pas standard du tout, et
%   que ce n'est pas si trivial que m=sf: il faut aussi faire zorn pour
%   compléter toute trace en une trace maximale, etc... donc là, dans
%   l'introduction c'est un peu délicat d'appuyer sur ces détails
%   relativement techniques. d'autant plus que c'est déjà un peu olé
%   olé, puisqu'on laisse le lecteur deviner les définitions standard à
%   partir des notres...
% }%
In contrast, our notion of play is more flexible than standard traces,
so that our counterpart to must testing equivalence actually
distinguishes these two processes: the infinite play where the output
prefix is not performed is not maximal, so that the corresponding
unfair behaviour is not taken into account. In other words, thanks to
our notion of play, the rather natural notion of must testing already
avoids what we call `spatial unfairness'.
However, must testing does not coincide with fair testing in our
setting, because there are other sources of unfairness, that are not
handled properly. Technically, we prove that $\bbot^m$ coincides with
the set of strategies whose states all admit a successful
extension. However, the restriction to finite plays in the definition
of $\bbot^f$ is required to rule out other sources of unfairness.
% we exhibit two processes, say $P$ and $Q$, which are
% equivalent for standard must testing, but whose translations
% $\transl{P}$ and $\transl{Q}$ are $\bbot^m$-equivalent.  \dam{c'est
%   dans l'autre sens, mais il faut reprendre de toutes façons}
% \tom{hurm \ldots} As a quick explanation: for the standard notion of
% trace used in the definition of must testing, infinite traces may
% never be extended; in our setting some infinite plays may be extended:
% our notion of play is more flexible than traces, as we explain in
% Section~\ref{subsec:fairmust}.

% \todo{Try to show that if we restrict to tests in the image of the
%   translation, then the semantics is fully abstract.}

\paragraph{Summary}
In summary, our approach emphasises a flexible notion of
multiple-player play, encompassing both views in the sense of game
semantics, closed-world plays, and intermediate notions.  Strategies
are then described as presheaves on plays, while innocent strategies
are presheaves on views. Innocent strategies admit a notion of
interaction, or amalgamation, and are embedded into strategies via
right Kan extension. This allows a notion of testing, or interactive
equivalence by amalgamation with the test, right Kan extension, and
finally restriction to closed-world.

Our main technical contributions are then a translation of CCS terms
with recursive equations into innocent strategies, and the study of
fair and must equivalences in our setting.

\paragraph{Perspectives}
Our next task is clearly to tighten the link with CCS. Namely, we
should explore which equivalence on CCS is induced via our
translation, for a given interactive equivalence. We will start with
$\bbot^f$.  Furthermore, the very notion of interactive equivalence
might deserve closer consideration. Its current form is rather
\emph{ad hoc}, and one could hope to see it emerge more naturally from
the game. For instance, the fixed class $\bbot$ of `successful'
strategies should probably be more constrained than is done here.
Also, the paradigm of observing via the set of successful tests might
admit sensible refinements, e.g., probabilistic ones.

Another possible research direction is to tighten the link with
`graphical' approaches to rewriting, such as graph rewriting or
computads.  E.g., our plays might be presented by a
computad~\cite{Guiraud}, or be the bicategory of rewrite sequences up
to shift equivalence, generated by a graph grammar in the sense of
Gadducci et al.~\cite{DBLP:journals/tcs/GadducciHL99}. Both goals
might require some technical adjustments, however.  For computads, we
would need the usual yoga of U-turns to flexibly model our positions;
e.g., zigzags of U-turns are usually only equal up to a
higher-dimensional cell, while they would map to equal positions in
our setting.  For graph rewriting, the problem is that our positions
are not exactly graphs (e.g., the channels known to a player are
linearly ordered).

Other perspectives include the treatment of more complicated calculi
like $\pi$ or $\lambda$. In particular, calculi with duplication of
terms will pose a serious challenge.  An even longer-term hope is to
be able to abstract over our approach. Is it possible to systematise
the process starting from a calculus as studied in programming
language theory, and generating its strategies modulo interactive
equivalence?  If this is ever understood, the next question is: when
does a translation between two such calculi preserve a given
interactive equivalence?  Finding general criteria for this might have
useful implications in programming languages, especially compilation.

\paragraph{Notation}
Throughout the paper, we abusively identify $n$ with $\ens{1 \ldots
  n}$, for readability. So, e.g., $i \in n$ means $i \in \ens{1,
  \ldots, n}$.

\begin{wrapfigure}{r}{0pt}
  \diag(.4,1){%
    |(FC)| FC \& |(FC')| FC' \\ %
    |(GD)| GD \& |(GD')| GD' %
  }{ %
    (FC) edge[labelu={F (f)}] (FC') %
    edge[labell={u}] (GD) %
    (FC') edge[labelr={u'}] (GD') %
    (GD) edge[labeld={G (g)}] (GD') %
    }
\end{wrapfigure}
The various categories and functors constructed in the development are
summed up with a short description in Table~\ref{tab:summary}. There,
given two functors $\C \xto{F} \E \xot{G} \D$, we denote (slightly
abusively) by $\C \downarrow_\E \D$ the \emph{comma} category: it has
as objects triples $(C,D,u)$ with $C \in \C$, $D \in \D$, and $u
\colon F (C) \to G (D)$ in $\E$, and as morphisms $(C,D,u) \to
(C',D',u')$ pairs $(f,g)$ making the square above commute. Also, when
$F$ is the identity on $\C$ and $G \colon 1 \to \C$ is an object $C$
of $\C$, this yields the usual \emph{slice} category, which we
abbreviate as $\C / C$. Finally, the category of presheaves on any
category $\C$ is denoted by $\Chat = [\op\C,\Set]$.

\begin{table}[t]
  \centering
  \begin{tabular}[t]{|c|c|}
    \hline
    Category & Description of its objects \\ \hline 
    \parbox[b][1.5em][b]{1cm}{$\Chat$} &  `diagrams'  \\
    $\B \into \Chat$ &  positions \\
    $\CE \into (\B \downarrow_{\Chat} \Chat) $ & plays \\
    $\CEX = (\CE \downarrow_{\B} (\B/X))$ & plays on a position $X$ \\
    $\CVX \into \CEX$ & views on $X$ \\
    $\SS_X = \CVXhat$ & innocent strategies on $X$ \\ 
    $\CW \into \CE$ & closed-world plays \\
    $\CWofX$ & closed-world plays on $X$ \\
    \hline
  \end{tabular}
  \caption{Summary of categories and functors}
  \label{tab:summary}
\end{table}

We denote by $\ob (\C)$ the set of objects of any small category $\C$.
For any functor $F \colon \C \to \D$, we denote by $\op\F \colon \op\C
\to \op \D$ the functor induced on opposite categories, defined
exactly as $F$ on both objects and morphisms. Also, recall that an
\emph{embedding} of categories is an injective-on-objects, faithful
functor. This admits the following generalisation: a functor $F \colon
\C \to \D$ is \emph{essentially injective on objects} when $FC \iso
FC'$ implies $C \iso C'$. Any faithful, essentially injective on
objects functor is called an \emph{essential embedding}.

\section{Plays as string diagrams}\label{sec:string}

We now describe our approach more precisely, starting with the
category of multiple-player plays. For the sake of clarity, we first
describe this category in an informal way, before giving the precise
definition (Section~\ref{sec:diagrams}).

\subsection{Positions}\label{subsec:positions}
\begin{wrapfigure}{r}{0pt}
      \diagramme[stringdiag={.6}{.6}]{}{%
        \path[-,draw] %
        (a) edge (j1) edge (j2) %
        (c) edge (j2) edge (j3) %
        (b) edge (j1) edge (j2) edge (j3) %
        ; %
        \node[diagnode,at= (j3.south east)] {\ \ \ } ; %
      }{%
        \& \canal{a}     \& \&  \canal{c} \\
    \joueur{j1} \& \& \joueur{j2} \& \& \joueur{j3} \\
    \& \& \canal{b} \&    
  }{%
  }%
\end{wrapfigure}
Since the game represents CCS, it should be natural that players are
related to each other via the knowledge of \emph{communication
  channels}. This is represented by a kind of\footnote{Only `a kind
  of', because, as mentioned above, the channels known to a player are
  linearly ordered.}  finite, bipartite graph: an example position is
on the right.  Bullets represent players, circles represent channels,
and edges indicate when a player knows a channel. The channels known
by a player are linearly ordered.  Formally, as explained in
Section~\ref{sec:diagrams}, positions are presheaves over a certain
category $\C_1$.  Morphisms of positions are natural transformations,
which are roughly morphisms of graphs, mapping players to players and
channels to channels. In full generality, morphisms thus do not have
to be injective, but include in particular embeddings of positions in
the intuitive sense. Positions and morphisms between them form a
category $\B$.
% Also notice that this
% representation allows to circumvent a syntactical difficulty in CCS,
% where the above position cannot be directly represented: the player in
% the center shares a private communication channel with each of the two
% outer players.
% \tom{c'est de moi hein, mais est-ce que 'starting from the lower' one
%   est pas un peu foireux? j'aurais maintenant tendance à préférer
%   'counterclockwise in $[-\pi/2, 3\pi/2[$'}
% \dam{$\pi$ c'est trop pointilleux, on pourrait carrément s'arrêter à
%   `linearly ordered'}

\subsection{Moves}

\emph{Plays} will be defined as glueings of \emph{moves} between
positions. Moves are derived from the very definition of CCS, as we
now sketch. The diagrams we draw in this section will be given a
very precise combinatorial definition in Section~\ref{sec:diagrams}.

%%%%%%%%% forking move

Let us start with the \emph{forking} move, which corresponds to
parallel composition in CCS: a process (the player) forks into two
sub-processes. In the case of a player knowing two channels, the
forking move is represented by the diagram
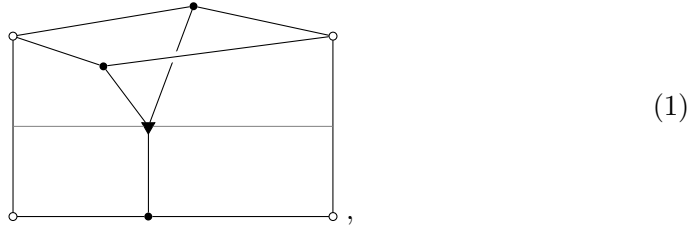
\begin{equation}
  \diagramme[stringdiag={.4}{.6}]{}{
%     \node[coordinate] (inter) at (intersection cs: %
%     first line={(t_1)-- (t1)}, %
%     second line={(para)-- (t_2)}) {} ; %
%     \path[-,dashed] (para) edge (inter) ; %
%     \path[-] (inter) edge (t_2) ; %
    \node[diagnode,at= (s1.south east)] {\ \ \ ,} ; %
  }{%
    \& \& \&  \& \joueur{t_2} \\
    \canal{t0} \& \& \& \& \& \& \& \canal{t1} \& \& \\ %\node[diagnode] (Y) {$Y$} ; \\
    \& \& \joueur{t_1}  \&  \\
    \& \ \& \\
    \coord{i0} \& \& \& \couppara{para} \& \& \& \& \coord{i1} \& \& \\ % \node[diagnode] (M) {$M$} ; \\
    \& \ \& \\
    \& \&  \\
    \canal{s0} \& \& \& \joueur{s} \& \& \& \& \canal{s1} \& \& %\node[diagnode] (X) {$X$} ; 
  }{%
    (para) edge[-] (t_1)
    (para) edge[-] (t_2) %
    (t0) edge[-] (t_1) %
    (t0) edge[-] (t_2) %
    (t1) edge[-,fore] (t_1) %
    (t1) edge[-] (t_2) %
    (s0) edge[-] (s) %
    (s1) edge[-] (s) %
    (s0) edge[-] (t0) %
    (s1) edge[-] (t1) %
    (s) edge[-] (para) %
    (i0) edge[-,gray,very thin] (para) %
    (i1) edge[-,gray,very thin] (para) %
%    (Y) edge[into] (M) %
%    (X)  edge[linto] (M) %
  }\label{eq:para}
\end{equation}
to be thought of as a move from the bottom position $X$
\begin{center}
  \diagramme[stringdiag={.4}{.6}]{baseline=(s.south)}{}{ %
    \canal{s0} \& \& \& \joueur{s} \& \& \& \& \canal{s1} %
  }{%
    (s) edge[-] (s1) edge[-] (s0) %
  }
\end{center}
(with one player $p$) to the top position $Y$
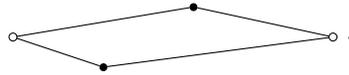
\begin{center}
  \diagramme[stringdiag={.4}{.6}]{}{
    \node[diagnode,at= (t1.south east)] {\ \ \ .} ; %
  }{%
    \& \& \&  \& \joueur{t_2} \\
    \canal{t0} \& \& \& \& \& \& \& \canal{t1} \& \& \\ %\node[diagnode] (Y) {$Y$} ; \\
    \& \& \joueur{t_1}  \&  \\
  }{%
    (t0) edge[-] (t_1) %
    (t0) edge[-] (t_2) %
    (t1) edge[-] (t_1) %
    (t1) edge[-] (t_2) %
  }
\end{center}
(with two players, which we call the `avatars' of $p$).  The left- and
right-hand borders are just channels evolving in time, not noticing
that the represented player forks into two. The surfaces spread
between those vertical lines represent links (edges in the involved
positions) evolving in time. For example, each link here divides into
two when the player forks, thus representing the fact that both of the
avatars retain knowledge of the corresponding channel.  There is of
course an instance $\paran$ of forking for each $n$, according to the
number of channels known to the player.  As for channels known to a
player, the players and channels touching the black triangle are
ordered: there are different `ports' for the initial player and its
two avatars.

%%%%%%%%% tick and channel creation moves

We then have a \emph{tick} move $\tickn$, whose role is to define
successful plays, and a move for the \emph{channel creation} or
\emph{restriction} of CCS, here $\nun$. In the case where the player
knows two channels, they are graphically represented as
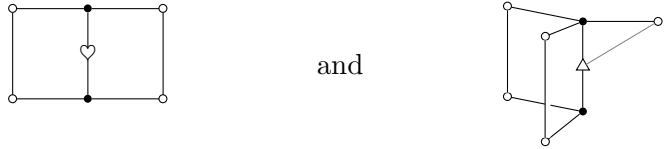
\begin{center}
  \diagramme[stringdiag={.6}{1}]{}{
    \path[-] (a) edge (a') %
    edge (p) %
    (tick) edge[shorten <=-1pt] (p) edge[shorten <=-1pt] (p') %
    (p') edge (a') edge (b') %
    (b) edge (p) edge (b') %
    ; %
%    \node[diagnode,at= (b.south east)] {\ \ \ .} ; %
  }{ %
    \canal{a'} \& \joueur{p'} \& \canal{b'} \\ %
    \& \couptick{tick} \& \\ %
    \canal{a} \& \joueur{p} \& \canal{b} \\
  }{%
  }
  \hfil and \hfil %
  \diagramme[stringdiag={.2}{.5}]{}{%
    \node[coordinate] (inter) at (intersection cs: %
    first line={(s)-- (s0)}, %
    second line={(s1)-- (t1)}) {} ; %
    \path[draw] (s) edge (inter) ; %
    \path[-,draw] %
    (s1) edge (s) %
    (t1) edge (t) %
    (t0) edge (t) %
    (t2) edge (t) %
    (t) edge (nu) %
    (s) edge (nu) %
    (s0) edge[fore] (t0) %
    (s1) edge[fore] (t1) %
    (nu) edge[gray,very thin] (t2) %
    (inter) edge (s0) %
    ; %
%    \node[diagnode,at= (s.base east)] {\ \ \ ,} ; %
  }{%
    \canal{t0}     \& \&  \\
    \& \& \joueur{t} \& \& \canal{t2} \& \\
    \& \canal{t1} \&    \\
    \\
    \& \& \coupnu{nu} \& \&  \& \\
    \\
    \canal{s0}     \& \&  \& \&  \\
    \& \& \joueur{s} \& \& \& \\\\
    \& \canal{s1} \&  \& \&  
  }{%
  }%
\end{center}
respectively. As expected, there is an instance of each of these two
moves for each number $n$ of channels known to the player.

%%%%%%%%% synchronisation move

We also need a move to model CCS-like synchronisation, between two
players.  For all $n$ and $m$, representing the numbers of channels
known to the players involved in the synchronisation, and for all $i
\in n$, $j \in m$, there is a \emph{synchronisation} $\taunimj$,
represented, in the case where one player outputs on channel $3 \in
3$ and the other inputs on channel $1 \in 2$, by
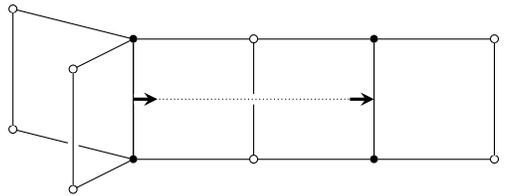
\begin{center}
  \diagramme[stringdiag={.4}{.8}]{}{%
    \node[coordinate] (inter) at (intersection cs: %
    first line={(s)-- (s0)}, %
    second line={(s1)-- (t1)}) {} ; %
    \path[draw] (s) edge (inter) ; %
      \path[-] %
      (s1) edge (s) %
      (s0) edge (inter) %
      (s2) edge (s) %
      (s2) edge (s') %
      (t1) edge (t) %
      (t0) edge (t) %
      (t2) edge (t) %
      (t2) edge (t') %
      (t) edge (iota.west) %
      (s) edge (iota.west) %
      (s') edge (iota'.east) %
      (t') edge (iota'.east) %
      (t'0) edge (t') %
      (s'0) edge (s') %
      (s'0) edge[fore] (t'0) %
      (s0) edge[fore] (t0) %
      (s1) edge[fore] (t1) %
      (s2) edge (t2) %
      ; %
    \path[-] %
    (iota) edge[fore,densely dotted] (iota') %
    ; %
    \node[diagnode,at= (s'0.south east)] {\ \ \ .} ; %
    \path (iota) --  (iota') node[coordinate,pos=.1] (iotatip) {} node[coordinate,pos=.9] (iotatip') {} ; %
      \path[-] (iota) edge[->,>=stealth,very thick] (iotatip) ; %
      \path[-] (iotatip') edge[->,>=stealth,very thick] (iota') ; %
      \foreach \x/\y in {s/t,s'/t'} \path[-] (\x) edge (\y) ; %
  }{%
    \canal{t0} \& \&  \&  \\
    \& \& \joueur{t} \& \& \canal{t2} \& \& %
    \joueur{t'} \& \&  \canal{t'0} \\
    \& \canal{t1} \&  \& \\
    \& \& \coupout{iota}{0} \& \&  \& \& \coupin{iota'}{0} \\
    \canal{s0} \& \&  \& \& \&  \& \\
    \& \& \joueur{s} \& \& \canal{s2} \& \& \joueur{s'} \& \& \canal{s'0} \\ \& \canal{s1} \& \& %
  }{%
  }%
\end{center}
% \dam{selon comment ça évolue, cette info peut attendre}%
As we shall see in Section~\ref{sec:diagrams}, the dotted wire in the
picture is actually a point in the formal representation (i.e., an
element of the corresponding presheaf).
% \tom{oui, surtout maintenant que les coups in et out isolés ont
%  déménagé plus bas}%

\medskip

The above four kinds of moves (forking, tick, channel creation, and
synchronisation) come from the reduction semantics of CCS. We classify
these as \emph{closed-world} moves, since they correspond to the
evolution of a group of players in isolation.

We however need a more fine-grained structure for moves: moves
whose final position has more than one player (forking and
synchronisation) must be decomposed into \emph{basic} moves, to get an
appropriate notion of view.

%%%%%%%%% half-forking moves

We introduce two sub-moves for forking: \emph{left} and \emph{right
  half-forking}. In the case where the player knows two channels, they are
represented by the following diagrams, respectively:
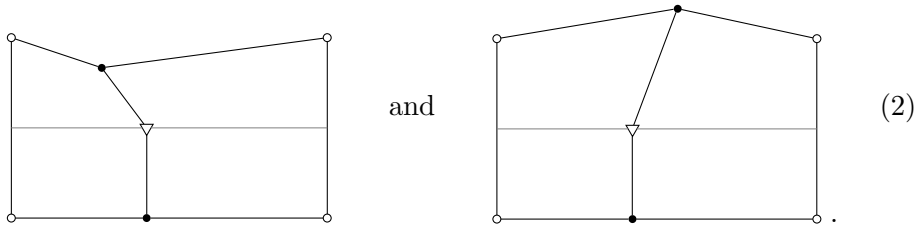
\begin{equation}
    \diagramme[stringdiag={.4}{.6}]{}{
  }{%
    \& \& \&  \&  \\
    \canal{t0} \& \& \& \& \& \& \& \canal{t1}  \\ %\node[diagnode] (Y) {$Y$} ; \\
    \& \& \joueur{t_1}  \&  \\
    \& \ \& \\
    \coord{i0} \& \& \& \coupparacreux{para} \& \& \& \& \coord{i1} \\ % \node[diagnode] (M) {$M$} ; \\
    \& \ \& \\
    \& \&  \\
    \canal{s0} \& \& \& \joueur{s} \& \& \& \& \canal{s1} %\node[diagnode] (X) {$X$} ; 
  }{%
    (t0) edge[-] (t_1) %
    (t1) edge[-] (t_1) %
    (s0) edge[-] (s) %
    (s1) edge[-] (s) %
    (s0) edge[-] (t0) %
    (s1) edge[-] (t1) %
    (s) edge[-] (para) %
    (para) edge[-] (t_1) %
    (i0) edge[-,gray,very thin] (para) %
    (i1) edge[-,gray,very thin] (para) %
%    (Y) edge[into] (M) %
%    (X)  edge[linto] (M) %
  }
\hspace*{.05\textwidth} \mbox{and} \hspace*{.05\textwidth}
    \diagramme[stringdiag={.4}{.6}]{}{
    \node[diagnode,at= (s1.south east)] {\ \ \ .} ; %
  }{%
    \& \& \&  \& \joueur{t_2} \\
    \canal{t0} \& \& \& \& \& \& \& \canal{t1} \\ %\node[diagnode] (Y) {$Y$} ; \\
    \& \&   \&  \\
    \& \ \& \\
    \coord{i0} \& \& \& \coupparacreux{para} \& \& \& \& \coord{i1} \\ % \node[diagnode] (M) {$M$} ; \\
    \& \ \& \\
    \& \&  \\
    \canal{s0} \& \& \& \joueur{s} \& \& \& \& \canal{s1} %\node[diagnode] (X) {$X$} ; 
  }{%
    (t0) edge[-] (t_2) %
    (t1) edge[-] (t_2) %
    (s0) edge[-] (s) %
    (s1) edge[-] (s) %
    (s0) edge[-] (t0) %
    (s1) edge[-] (t1) %
    (s) edge[-] (para) %
    (para) edge[-] (t_2) %
    (i0) edge[-,gray,very thin] (para) %
    (i1) edge[-,gray,very thin] (para) %
  }\label{eq:paraviews}
\end{equation}
These sub-moves represent what each of the `avatars' of the forking
player sees of the move. We call $\paraln$ and $\pararn$ the
respective instances of the left-hand and right-hand basic moves for a
player knowing $n$ channels. Formally, there will be injections from the
left and right half-forking moves to the corresponding forking moves.

%%%%%%%%% input/output moves

We finally decompose synchronisation into an input move and an output
move: $a.P$ and $\abar.P$ in CCS become $\iotaposni$ and $\iotanegni$
here (where $n$ is the number of known channels, $i \in \ens{1 \ldots
  n}$ is the index of the channel bearing the synchronisation).  Here,
output on the right-hand channel and input on the left-hand channel
respectively look like
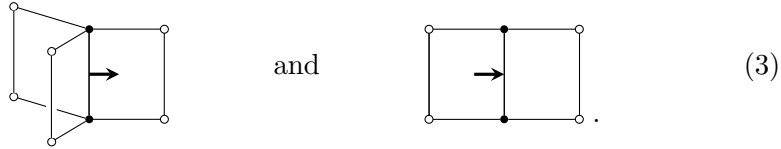
\begin{equation}
    \diagramme[stringdiag={.3}{.5}]{baseline=($(iota.south)$)}{%
    \node[coordinate] (inter) at (intersection cs: %
    first line={(s)-- (s0)}, %
    second line={(s1)-- (t1)}) {} ; %
    \path[draw] (s) edge (inter) ; %
      \path[-] %
      (s0) edge (inter) %
      (s2) edge (s) %
      (t1) edge (t) %
      (t0) edge (t) %
      (t2) edge (t) %
      (t) edge (iota.west) %
      (s) edge (iota.west) %
      (s0) edge[fore] (t0) %
      (s1) edge[fore] (t1) %
      (s2) edge (t2) %
      (s1) edge (s) %
      ; %
      \moveout{iota}{0} %
      \foreach \x/\y in {s/t} \path[-] (\x) edge (\y) ; %
  }{%
    \canal{t0} \&  \\
    \& \& \joueur{t} \& \& \canal{t2} \\
    \& \canal{t1} \\
    \& \& \coupout{iota}{0}  \\
    \canal{s0} \& \&   \\
    \& \& \joueur{s} \& \& \canal{s2}  \\
    \& \canal{s1} \& \& %
  }{%
  }%
  \hspace*{.1\textwidth} \mbox{and}   \hspace*{.1\textwidth}
  \diagramme[stringdiag={.6}{1}]{baseline=($(in.south)$)}{
    \path[-] (a) edge (a') %
    edge (p) %
    (in.west) edge (p) edge (p') %
    (p') edge (a') edge (b') %
    (b) edge (p) edge (b') %
    ; %
    \node[diagnode,at= (b.south east)] {\ \ \ .} ; %
    \movein{in}{180} %
    \foreach \x/\y in {p/p',a/a'} \path[-] (\x) edge (\y) ; %
  }{ %
    \canal{a'} \& \joueur{p'} \& \canal{b'} \\ %
    \coord{bin} \& \coupout{in}{0} \& \\ %
    \canal{a} \& \joueur{p} \& \canal{b} \\
  }{%
  } %
  \label{eq:inout}
\end{equation}
Like with forking, there will be injections from the input and output
moves to the corresponding synchronisation moves.

\medskip

All in all, there are three classes of moves, which we summarise in
Table~\ref{tab:moves}.

\begin{itemize}
\item Tick, channel creation, half-forking, and input/output moves are
  \emph{basic} moves: they evolve from a position with exactly one
  player to another position with exactly one player. These moves are
  used to define views later on.
\item Forking, synchronisation, tick and channel creation moves are
  \emph{closed-world} moves: they correspond to the case where a group
  of players evolves on its own, in isolation; they are central to the
  notion of interactive equivalence.
\item We need a third class of moves, called \emph{full}, which
  consists of forking, input, output, tick and channel creation. They
  involve a single player and all of its avatars. They appear, e.g.,
  in the statement of Lemma~\ref{lem:jointsmaps}, which is a partial
  correctness criterion for closed-world plays.
\end{itemize}

\begin{table}[t]
  \centering
  \begin{tabular}[t]{c|c|c}
    Basic & Full & Closed-world \\  \hline & &  \\[-.5em]
    \begin{minipage}[c]{0.3\linewidth} \centering
      Left half-forking \\
      Right half-forking 
    \end{minipage}
    & 
    Forking & Forking \\[1em]
    \begin{minipage}[c]{0.2\linewidth} \centering
      Input \\
      Output
    \end{minipage}
    & 
    \begin{minipage}[c]{0.2\linewidth} \centering
      Input \\
      Output
    \end{minipage}
    & 
    Synchronisation \\[1em]
    Channel creation & Channel creation & Channel creation \\[1em]
    Tick & Tick & Tick 
  \end{tabular}
  \caption{Summary of classes of moves}
  \label{tab:moves}
\end{table}

\noindent
Formally, we define moves as cospans $X \into P \otni Y$ in the
category of diagrams (technically a presheaf category $\Chat$---see
Section~\ref{sec:diagrams}), where $X$ is the initial position and
$Y$ the final one.  Both legs of the cospan are actually monic morphisms
in $\Chat$, as will be the case for all cospans considered here.
%\tom{pour `starting' et `ending', j'ai mis un peu partout du `initial'
%  and `final'. les deux choix me vont mais on uniformise non? tu
%  préfères quoi?} %
%\dam{ok pour initial et final} %
% In the diagrams which we depicted for the various moves, the (more or
% less) vertical lines represent dots (channels and players) moving in
% time, upwards.  %
% \tom{La dernière phrase redonde un peu avec les
%   explications du début, on pourrait la virer non?} %

\subsection{Plays}
\begin{figure}[t]
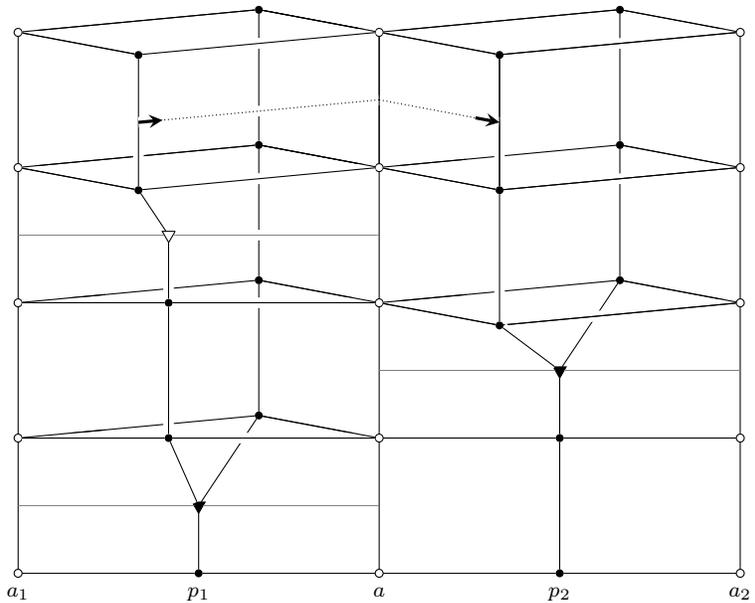

  \begin{center}
    \diagramme[stringdiag={.3}{.4}]{}{ \node[diagnode,at= (a2.south
      east)] {\ \ \ .} ; %
      \foreach \x/\y in {a1/a1',a1'/a1'',a1''/a1''',a1'''/a1'''',a2/a2',a2'/a2'',a2''/a2''',a2'''/a2'''',a/a',a'/a'',a''/a''',a'''/a'''',%
        p1/parap1,parap1/p12,p12/p12',p12'/p12'',a1/p1,p1/a,a1'/p12,p12/a',a1''/p12',p12'/a'',a/p2,p2/a2,a'/p2',%
        p2/p2',p2'/a2',p2'/parap2,parap2/p22,a''/p22,p22/a2'',a'''/p22',p22'/a2''',p12''/p12''',p22/p22',p22'/p22'',%
        a''''/p22'',p22''/a2''''%
      } 
      \path[-] (\x) edge (\y) ; %
      \foreach \a/\p/\b in {a1'''/p12''/a''',%a1'''/p112/a''',
        a1'''/p111/a''',a1''''/p12'''/a'''',%a1''''/p112'/a'''',
        a1''''/p111'/a'''',%
        a1'/p11/a',a1''/p11'/a'',a''/p21/a2'',a'''/p21'/a2''',a''''/p21''/a2''''} 
      \path[-]  (\p) edge[fore] (\a) edge[fore] (\b) %
      ; %
      \foreach \a/\p/\b in {a1'''/p12''/a''',%a1'''/p112/a''',
        a1'''/p111/a''',a1''''/p12'''/a'''',%a1''''/p112'/a'''',
        a1''''/p111'/a'''',%
        a1'/p11/a',a1''/p11'/a'',a''/p21/a2'',a'''/p21'/a2''',a''''/p21''/a2''''} 
      \path[-]  (\p) edge (\a) edge (\b) %
      ; %
      \foreach \x/\y in {parap1/p11,p11/p11',p11'/parap11,%parap11/p112,parap11/p111,
        p111/a''',%
      %p112/p112',p111/p111',
        p111'/a'''',parap2/p21,p21/p21',p21'/p21'',p111/p111'} %
      \path[-] (\x) edge[fore] (\y) %
      ; %
      \foreach \x/\y in {%p112/a''',
        p12/a',p12'/a'',p12''/a''',p22/a2'',p22'/a2''',p22''/a2'''',%p112'/a'''',
        p12'''/a'''',p22''/a''''} %
      \path[-] (\x) edge (\y) ; %
      \foreach \x/\y in {p22/a'',p22'/a''',p12/a1',p12'/a1''} %
      \path[-] (\x) edge[shorten <=2cm] (\y) ; %
      \foreach \para/\a/\ai in {parap1/a1parap1/aparap1,
        parap2/a2parap2/aparap2, parap11/a1parap11/aparap11}
      \path[draw] (\para) edge[-,gray,very thin] (\a) %
      edge[-,gray,very thin] (\ai) %
      ; %
      \path[-] (parap11) edge[-] (p111) ; %
      \path[-] (parap11) edge[-,fore,gray,very thin,shorten <=1pt, shorten >=1pt] (aparap11) ; %
      \path[-] (out11) edge[fore,densely dotted] (aout11) %
      (aout11) edge[fore,densely dotted] (in11) %
      ; %
      \path (out11) --  (aout11) node[coordinate,pos=.1] (out11tip) {} ; %
      \path (in11) -- (aout11) node[coordinate,pos=.2] (in11tip) {} ; %
      \path[-] (out11) edge[->,>=stealth,very thick] (out11tip) ; %
      \path[-] (in11tip) edge[->,>=stealth,very thick] (in11) ; %
      \foreach \x/\y in {a'''/a'''',p21'/p21''} \path[-] (\x) edge (\y) ; %
      \node[at=(p1.south),anchor=north] {$\scriptstyle p_1$} ; %
      \node[at=(p2.south),anchor=north] {$\scriptstyle p_2$} ; %
      \node[at=(a.south),anchor=north] {$\scriptstyle a$} ; %
      \node[at=(a1.south),anchor=north] {$\scriptstyle a_1$} ; %
      \node[at=(a2.south),anchor=north] {$\scriptstyle a_2$} ; %
    }{%
      \& \& \& \& \& \& \& \& \joueur{p12'''} \& \& \& \& \& \& \& \& \& \& \& \& \joueur{p22''} \& \& \\
      \canal{a1''''} \& \& \& \& \& \& % \joueur{p112'}
      \& \& \& \& \& \& \canal{a''''} \& \& \& \& \& \& \& \& \& \& \& \& \canal{a2''''} \\
      \& \& \& \& \joueur{p111'} \& \& \& \& \& \& \& \& \& \& \& \& \joueur{p21''} \& \& \\
      \& \& \\
      \coord{a1out11} \& \& \& \&  \&  \& \& \& \& \& \& \& \coord{aout11} \& \& \& \& \& \& \\
      \coord{a1out11} \& \& \& \& \coupout{out11}{4} \&  \& \& \& \& \& \& \&  \& \& \&  \& \coupin{in11}{-7} \& \& \\
      \& \& \& \& \& \& \& \& \joueur{p12''} \& \& \& \& \& \& \& \& \& \& \& \& \joueur{p22'} \& \& \\
      \canal{a1'''} \& \& \& \& \& \& % \joueur{p112}
      \& \& \& \& \& \& \canal{a'''} \& \& \& \& \& \& \& \& \& \& \& \& \canal{a2'''} \\
      \& \& \& \& \joueur{p111} \& \& \& \& \& \& \& \& \& \& \& \& \joueur{p21'} \& \& \\
      \& \& \\
      \coord{a1parap11} \& \& \& \& \& \coupparacreux{parap11} \& \& \& \& \& \& \& \coord{aparap11} \& \& \& \& \& \& \\
      \& \& \\
      \& \& \& \& \& \& \& \& \joueur{p12'} \& \& \& \& \& \& \& \& \& \& \& \& \joueur{p22} \& \& \\
      \canal{a1''} \& \& \& \& \& \joueur{p11'} \& \& \& \& \& \& \& \canal{a''} \& \& \& \& \& \& \& \& \& \& \& \& \canal{a2''} \\
      \& \& \& \&  \& \& \& \& \& \& \& \& \& \& \& \& \joueur{p21} \& \& \\
      \& \& \\
      \coord{a1parap2} \& \& \& \& \&  \& \& \& \& \& \& \& \coord{aparap2} \& \& \& \& \& \& \couppara{parap2} \& \& \& \& \& \& \coord{a2parap2} \\
      \& \& \\
      \& \& \& \& \& \& \& \& \joueur{p12} \& \& \& \& \& \& \& \& \& \& \& \&  \& \& \\
      \canal{a1'} \& \& \& \& \& \joueur{p11} \& \& \& \& \& \& \& \canal{a'} \& \& \& \& \& \& \joueur{p2'} \& \& \& \& \& \& \canal{a2'} \\
      \& \& \& \&  \& \& \& \& \& \& \& \& \& \& \& \&  \& \& \\
      \& \& \\
      \coord{a1parap1} \& \& \& \& \&  \& \couppara{parap1} \& \& \& \& \& \& \coord{aparap1} \& \& \& \& \& \&  \& \& \& \& \& \&  \\
      \& \& \\
      \& \& \& \& \& \& \& \&  \& \& \& \& \& \& \& \& \& \& \& \&  \& \& \\
      \canal{a1} \& \& \& \& \&  \& \joueur{p1} \& \& \& \& \& \& \canal{a} \& \& \& \& \& \& \joueur{p2} \& \& \& \& \& \& \canal{a2} \\
    }{%
    }
  \end{center}
  \caption{An example play}
\label{fig:exex}
\end{figure}

We now sketch how plays are defined as glueings of moves. We start
with the following example, depicted in Figure~\ref{fig:exex}. The
initial position consists of two players $p_1$ and $p_2$ sharing
knowledge of a channel $a$, each of them knowing another channel, resp.\
$a_1$ and $a_2$.  The play consists of four moves: first $p_1$ forks
into $p_{1,1}$ and $p_{1,2}$, then $p_2$ forks into $p_{2,1}$ and
$p_{2,2}$, and then $p_{1,1}$ does a left half-fork into $p_{1,1,1}$;
finally $p_{1,1,1}$ synchronises (as the sender) with $p_{2,1}$. Now,
we reach the limits of the graphical representation, but the order in
which $p_1$ and $p_2$ fork is irrelevant: if $p_2$ forks before $p_1$,
we obtain the same play. This means that glueing the various parts of
the picture in Figure~\ref{fig:exex} in different orders formally
yields the same result (although there are subtle issues in
representing this result graphically in a canonical way).

Let us now sketch a definition of plays. Recall that moves may be seen
as cospans $X \into M \otni Y$, and consider an \emph{extended} notion
of move, which may occur in a position not limited to players involved
in the move.  For example, the moves in Figure~\ref{fig:exex} are
extended moves in this sense.
%Let also a \emph{restriction} from $X$ to $Y$ be a cospan $X
%\xinto{\id} X \otni Y$, where $X$ and $Y$ are positions and $Y \into
%X$ is monic.  
  \begin{defn}\label{def:play}
    A \emph{play} is an embedding $X_0 \into U$ in the category
    $\Chat$ of diagrams, isomorphic to a possibly denumerable
    `composition' of moves %and restrictions
    in the (bi)category $\Cospan{\Chat}$ of cospans in $\Chat$, i.e.,
    obtained as a colimit:
      \begin{center}
        \diagramme[diagorigins={.7}{1.2}]{}{%
          \pbk{X_0}{U}{X_nnn} %
        }{%
          |(X_0)| X_0 \& \& |(X_1)| X_1 \& \ldots \& |(X_n)| X_n \&  \& |(X_nn)| X_{n+1} \& \& |(X_nnn)| X_{n+2} \& \ldots \\
          \& |(M_0)| M_0 \& \& \ldots \& \& |(M_n)| M_n \& \& |(M_nn)| M_{n+1} \& \ldots \\
          \& {\ } \\
          \& \& \& \& |(U)| U, %
        }{%
          (X_0) edge[into] (M_0) %
%          edge[into] (U) %
          (X_1) edge[linto] (M_0) %
%          edge[into] (U) %
          (X_n) edge[into] (M_n) %
%          edge[into] (U) %
          (X_nn) edge[linto] (M_n) %
%          edge[into] (U) %
          (X_nn) edge[into] (M_nn) %
          (X_nnn) edge[linto] (M_nn) %
%          edge[into] (U) %
          (M_0) edge[into] (U) %
          (M_n) edge[linto] (U) %
          (M_nn) edge[linto] (U) %
        }
      \end{center}
      where each $X_i \into M_i \otni X_{i+1}$ is %either
      an extended move.% or a restriction.
    \end{defn}
    We often denote plays just by $U$, leaving the embedding $X \into
    U$ implicit.
    \begin{rk}
      For finite plays, one might want to keep track not only of the
      initial position, but also of the final position. This indeed
      makes sense. Finite plays then compose `vertically', and form a
      double category. But infinite plays do not really have any final
      position, which explains our definition.
    \end{rk}
  
\begin{wrapfigure}{r}{0pt}
      \diag(.6,.8){%
        |(U)| U \& |(V)| V \\
        |(X)| X \& |(Y)| Y. %
      }{%
        (X) edge[into] (U) edge[labelu={h}] (Y) %
        (U) edge[labelu={k}] (V) %
        (Y) edge[into] (V) %
      }
\end{wrapfigure}
Let a morphism $(X \into U) \to (Y \into V)$ of plays be a pair
$(h,k)$ making the diagram on the right commute in $\Chat$.  This
permits both inclusion `in width' and `in height'. E.g., the play
consisting of the left-hand basic move in~\eqref{eq:paraviews} embeds
in exactly two ways into the play of Figure~\ref{fig:exex}. (Only two
because the image of the base position must lie in the base position
of the codomain.)  We have:
\begin{prop}\label{prop:cat}
  Plays and morphisms between them form a category $\E$.
\end{prop}
There is a projection functor $\E \to \B$ mapping each play $X \into
U$ to its base position $X$.  This functor has a section, which is an
embedding $\B \into \E$, mapping each position $X$ to the `identity'
play $X \into X$ on $X$.

\begin{rk}[Size]
  The category $\CE$ is only locally small. Since presheaves on a
  locally small category are less well-behaved than on a small
  category, we will actually consider a skeleton of $\CE$.
  Because $\CE$ consists only of denumerable presheaves, this skeleton
  is a small category. Thus, our presheaves in the next section may be
  understood as taken on a small category.
\end{rk}

\begin{rk}
  Plays are not very far from being just (infinite) abstract syntax
  trees (or forests) `glued together along channels'. E.g., the play
  from Figure~\ref{fig:exex} is the glueing of, say $(\paralof{2}
  (\abar.0)) | 0$ and $a | 0$ along $a$. 
\end{rk}

\subsection{Relativisation}\label{subsec:relativisation}
If we now want to restrict to plays over a given base position $X$, we may consider
\begin{defn}
  Let the category $\CEX$ have
  \begin{itemize}
  \item as objects pairs of a play $Y \into U$ and a morphism $Y \to X$,
  \item as morphisms $(Y \into U) \to (Y' \into U')$ all pairs $(h,k)$
    making the diagram
    \begin{center}
  \diag(.25,1.5){%
    |(U)| U \& \& |(U')| U' \\
    \\
    |(Y)| Y \& \& |(Y')| Y' \\
    \& |(X)| X %
  }{%
    (U) edge[labelu={k}] (U') %
    (Y) edge[into] (U) %
    edge [labelu={h}] (Y') %
    edge (X)
    (Y') edge[into] (U') %
    edge (X)
  }%
    \end{center}
    commute in $\Chat$.
  \end{itemize}
\end{defn}

We will usually abbreviate $U \otni Y \to X$ as just $U$ when no
ambiguity arises.  As for morphisms of positions, in full generality,
$h$ and $k$, as well as the morphisms $Y \to X$, do not have to be
injective. 
\begin{ex}
  Let $X$ be the position
  \diagramme[stringdiag={.1}{.2}]{baseline=(s_2.south)}{}{ %
    \canal{s0} \& \& \& \joueur{s_1} \& \& \& \& %
    \canal{s1} \& \& \& \joueur{s_2} \& \& \& \& %
    \canal{s2} \& \& \& \joueur{s_3} \& \& \& \& %
    \canal{s3} %
  }{%
    (s_1) edge[-] (s1) edge[-] (s0) %
    (s_2) edge[-] (s1) edge[-] (s2) %
    (s_3) edge[-] (s2) edge[-] (s3) %
  }.  %
  The play in Figure~\ref{fig:exex}, say $Y \into U$, equipped with
  the injection $Y \into X$ mapping the two players of $Y$ to the two
  leftmost players of $X$, is an object of $\CEX$.

  One naively could imagine that the objects $\CEX$ could just consist
  of plays $X \into U$ on $X$. However, spatial decomposition,
  Theorem~\ref{thm:spatial}, relies on our slightly more complex
  definition. E.g., still in Figure~\ref{fig:exex}, this allows us to
  distinguish between the identity view $[2] \idto [2] \xinto{p_1} X$
  on $p_1$ from the identity view $[2] \idto [2] \xinto{p_2} X$ on
  $p_2$, which would otherwise not be possible.
\end{ex}

\section{Diagrams}\label{sec:diagrams}
In this section, we define the category on which the string diagrams
of the previous section are presheaves. The techniques used here date
back at least to Carboni and
Johnstone~\cite{DBLP:journals/mscs/CarboniJ95,DBLP:journals/mscs/Johnstone04}.

\subsection{First steps}
Let us first consider two small examples. It is well-known that
directed graphs form a presheaf category: consider the category $\C$
freely generated by the graph with two vertices, say $\star$ and
$[1]$, and two edges $d, c \colon \star \to [1]$ between them.  One
way to visualise this is to compute the \emph{category of elements} of
a few presheaves on $\C$. Recall that the category of elements of a
presheaf $F$ on $\C$ is the comma category $y \downarrow_{\Chat} F$,
where $y$ is the Yoneda embedding. Via Yoneda, it has as elements
pairs $(C, x)$ with $C \in \ob (\C)$ and $x \in F (C)$, and morphisms
$(C,x) \to (D,y)$ morphisms $f \colon C \to D$ in $\C$ such that $F
(f) (y) = x$ (which we abbreviate as $y \cdot f = x$ when the context
is clear).
\begin{ex}
  Consider the graph
  \begin{center}
    \diag{%
      |(0)| 0 \& |(1)| 1 \& |(2)| 2 %
    }{%
      (0) edge[labelu={e}] (1) %
      (1) edge[labelu={e'}] (2) %
    }
  \end{center}
  with three vertices $0,1$, and $2$, and two edges $e$ and $e'$.

  This graph is represented by the presheaf $F$ defined by the
  following equations, whose category of elements is actually freely
  generated by the graph on the right:
\begin{center}
    \begin{minipage}[c]{0.3\linewidth}
      \begin{itemize}
      \item $F (\star) = \ens{0,1,2}$,
      \item $F ([1]) = \ens{e, e'}$,
      \end{itemize}
    \end{minipage}
    \hfil
    \begin{minipage}[c]{0.2\linewidth}
      \begin{itemize}
      \item $e \cdot d = 0$,
      \item $e \cdot c = 1$,
      \item $e' \cdot d = 1$,
      \item $e' \cdot c = 2$,
      \end{itemize}
    \end{minipage}
      \hfil
    \diag(.15,.8){%
      \& \& |(un)| 1 \& \\ %
      \& |(e)| e \& \& |(e')| e' \\ %
      |(zer)| 0 \& \& \& \& |(deux)| 2. %
    }{%
      (zer) edge[labelal={d}] (e) %
      (un) edge[labelal={c}] (e) %
      edge[labelar={d}] (e') %
      (deux) edge[labelar={c}] (e') %
    }
\end{center}
This latter graph is not exactly the
original one, but it does represent it. Indeed, for each vertex we
know whether it is in $F (\star)$ or $F ([1])$, hence whether it
represents a `vertex' or an `edge'. The arrows all go from a `vertex'
$v$ to an `edge' $e$. They lie over $d$ when $v$ is the domain of
$e$, and over $c$ when $v$ is the codomain of $e$.
\end{ex}

Multigraphs, i.e., graphs whose edges have a list of sources instead
of just one, may also be seen as a presheaves on the category freely
generated by the graph with
\begin{itemize}
\item as vertices: one special vertex $\star$, plus for each natural
  number $n$ a vertex, say, $[n]$; and
\item for all $n \in \Nat$, $n+1$ edges $\star \to [n]$, called $d_1, \ldots, d_n$, and $c$.
\end{itemize}
It should be natural for presheaves on this category to look like
multigraphs: the elements of a presheaf $F$ over $\star$ are the
vertices in the multigraph, the elements over $[n]$ are the $n$-ary
multiedges, and the action of the $d_i$'s give the $i$th source of a
multiedge, while the action of $c$ gives its target.
\begin{ex}
  Similarly, computing a few categories of elements might help
  visualising. As above, consider $F$ defined by
\begin{center}
    \begin{minipage}[c]{0.4\linewidth}
      \begin{itemize}
      \item $F (\star) = \ens{0,1,2,3,4,5}$,
      \item $F ([1]) = F ([0]) = \emptyset$,
      \item $F ([2]) = \ens{e'}$,
      \item $F ([3]) = \ens{e}$,
      \item $F ([n+4]) = \emptyset$,
      \end{itemize}
    \end{minipage}
    \hfil
    \begin{minipage}[c]{0.2\linewidth}
      \begin{itemize}
      \item $e \cdot c = 0$,
      \item $e \cdot d_1 = 1$,
      \item $e \cdot d_2 = 2$,
      \item $e \cdot d_3 = 3$,
      \end{itemize}
    \end{minipage}
    \hfil
    \begin{minipage}[c]{0.3\linewidth}
      \begin{itemize}
      \item $e' \cdot c = 1$,
      \item $e' \cdot d_1 = 4$,
      \item $e' \cdot d_2 = 5$,
      \end{itemize}
    \end{minipage}
\end{center}
whose category of elements is freely generated by the graph:
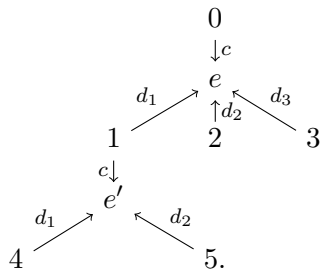
\begin{center}
    \diag(.3,.8){%
      \& \& |(zero)| 0 \& \\ %
      \& \& |(e)| e \& \\ %
      \& |(un)| 1 \& |(deux)| 2 \& |(trois)| 3 \\ %
      \& |(e')| e' \& \\ %
      |(quatre)| 4 \& \& |(cinq)| 5. %
    }{%
      (zero) edge[labelr={c}] (e) %
      (un) edge[labelal={d_1}] (e) %
      edge[labell={c}] (e') %
      (deux) edge[labelr={d_2}] (e) %
      (trois) edge[labelar={d_3}] (e) %
      (quatre) edge[labelal={d_1}] (e') %
      (cinq) edge[labelar={d_2}] (e') %
    }
\end{center}
\end{ex}

Now, this pattern may be extended to higher dimensions. Consider for
example extending the previous base graph with a vertex $[m_1,
\ldots,m_n; p]$ for all natural numbers $n, p, m_1, \ldots, m_n$, plus
edges
  $$
  \begin{array}{l}
    s_1\colon [m_1] \to [m_1, \ldots, m_n; p], \\
    \ldots, \\
    s_n\colon [m_n] \to [m_1, \ldots, m_n;p], and \\
    t \colon [p] \to [m_1, \ldots,
    m_n;p].
\end{array}
$$
Let now $\C$ be the free category on this extended
graph. Presheaves on $\C$ are a kind of 2-multigraphs: they have
vertices, multiedges, and multiedges between multiedges.

We could continue this in higher dimensions.

\subsection{Constructing the base category}
Our base category follows a very similar pattern. We start from a
slightly different graph: let $\G_0$ have just one vertex $\star$; let
$\G_1$, have one vertex $\star$, plus a vertex $[n]$ for each natural
number $n$, plus $n$ edges $d_1, \ldots, d_n \colon \star \to [n]$.
Let $\C_0$ and $\C_1$ be the categories freely generated by $\G_0$ and
$\G_1$, respectively. So, presheaves on $\C_1$ are a kind of 
%\dam{undirected?}
hypergraphs with arity (since vertices incident to a hyperedge are
numbered). This is enough to model positions.
\begin{ex}
  The position drawn at the beginning of
  Section~\ref{subsec:positions} may be represented as the presheaf
\begin{center}
    \begin{minipage}[c]{0.3\linewidth}
      \begin{itemize}
      \item $\star \mapsto \ens{1,2,3}$,
      \item $[2] \mapsto \ens{x,z}$,
      \item $[3] \mapsto \ens{y}$,
      \item $\_ \mapsto \emptyset$,
      \end{itemize}
    \end{minipage}
    \hfil
    \begin{minipage}[c]{0.2\linewidth}
      \begin{itemize}
      \item $x \cdot d_1 = 1$,
      \item $x \cdot d_2 = 2$,
      \item $z \cdot d_1 = 2$,
      \item $z \cdot d_2 = 3$,
      \end{itemize}
    \end{minipage}
    \hfil
    \begin{minipage}[c]{0.2\linewidth}
      \begin{itemize}
      \item $y \cdot d_1 = 1$,
      \item $y \cdot d_2 = 2$,
      \item $y \cdot d_3 = 3$,
      \end{itemize}
    \end{minipage}
\end{center}
whose category of elements is:
      \begin{center}
        \Diag(.6,.6){%
          \path[->,draw] %
          (a) edge[labelal={d_1}] (j1) edge[labelar={d_1}] (j2) %
          (c) edge[labelal={d_3}] (j2) edge[labelar={d_2}] (j3) %
          (b) edge[labelbl={d_2}] (j1) edge[labell={d_2}] (j2)
          edge[labelbr={d_1}] (j3) %
          ; %
          \node[diagnode,at= (j3.south east)] {\ \ \ } ; %
        }{%
          \& |(a)| 1     \& \&  |(c)| 3 \\
          |(j1)| x \& \& |(j2)| y \& \& |(j3)| z \\
          \& \& |(b)| 2. \& }{%
        }%
      \end{center}
\end{ex}

Now, consider the graph $\G_2$, which is $\G_1$ augmented with:
\begin{itemize}
\item for all $n$, vertices $\tickn$, $\paraln$, $\pararn$, $\nun$, 
\item for all $n$ and $1 \leq i \leq n$, vertices $\iotaposni$ and $\iotanegni$,
\item for all $n$, edges $s, t \colon [n] \to \tickn$, $s, t \colon [n] \to \paraln$, $s, t \colon [n] \to \pararn$, 
$s \colon [n] \to \nun$, $t \colon [n+1] \to \nun$, 
\item for all $n$ and $1 \leq i \leq n$, edges $s, t \colon [n] \to \iotaposni$, $s,t \colon [n] \to \iotanegni$.
\end{itemize}
We slightly abuse language here by calling all these $t$'s
and $s$'s the same. We could label them with their codomain, but we
refrain from doing so for the sake of readability.

Now, let $\C_2$ be the category generated by $\G_2$ and the relations
$s \rond d_i = t \rond d_i$ for all $n$ and $1 \leq i \leq n$ (for all
sensible---common---codomains). The intuition here is that for any
basic move by a player with $n$ channels, these $n$ channels remain
the same after the move. This includes the case of $\nun$, for which
the absence of any equation involving the new channel makes it
different from the others.

\begin{ex}\label{ex:in}
  Again, computing a few categories of elements is in order. For
  example, the category of elements of (the representable presheaf on)
  $\iotaof{3}{3}^-$ is the poset freely generated by the graph
  \begin{center}
    \diag(\diaghautici,.6){%
      |(t1)| t  d_1     \& \&  \& \&  \\
      \& \& |(t)| t \& \& \& \& |(t3)| t  d_3 \\
      \& |(t2)| t  d_2 \&  \& \&  \\
      \\
      \& \& |(iota)| \id_{\iotaof{3}{3}^-} \& \&  \\
      \\
      |(s1)| s  d_1     \& \&  \& \&  \\
      \& \& |(s)| s \& \& \& \& |(s3)| s  d_3 \\
      \& |(s2)| s d_2, \& \& \& }{%
      (s2) edge (s) %
      (s1) edge (s) %
      (s3) edge (s) %
      (t2) edge (t) %
      (t1) edge (t) %
      (t3) edge (t) %
      (t) edge (iota) %
      (s) edge (iota) %
      (s1) edge[fore,identity] (t1) %
      (s2) edge[fore,identity] (t2) %
      (s3) edge[fore,identity] (t3) %
    }%
  \end{center}
  to be compared with the corresponding pictures~\eqref{eq:inout}.
\end{ex}

\begin{ex}
  Similarly, the category of elements of $\nuof{1}$ is the poset
  freely generated by the graph
  \begin{center}
    \diag{%
      |(t1)| t d_1 \& |(t)| t \& |(t2)| t d_2 \\
      \& |(nu)| \id_{\nuof{1}} \& \\
      |(s1)| s d_1 \& |(s)| s.  }{%
      (t1) edge (t) %
      (t2) edge (t) %
      edge (nu) %
      (t) edge (nu) %
      (s1) edge (s) %
      (s) edge (nu) %
      (t1) edge[identity] (s1) %
    }
  \end{center}
Note that only channel creation changes the number of channels known to the
player, and accordingly the corresponding morphism $t$ has domain
$[n+1]$. 
\end{ex}

Presheaves on $\C_2$ are enough to model basic moves, but since we
want more, we continue, as follows.

Let $\G_3$ be $\G_2$, augmented with:
\begin{itemize}
\item for all $n$, a vertex $\paran$, and 
\item edges $l \colon \paraln \to \paran$ and $r \colon \pararn \to \paran$.
\end{itemize}
\begin{defn}\label{def:C3}
  Let $\C_3$ be the category generated by $\G_3$, the previous
  relations, plus the relations $l \rond s = r \rond s$.
\end{defn}
The equation models the fact that a forking
move should be played by just one player.  We also call $s = l \rond s
= r \rond s$ the common composite, which gives a uniform notation for
the initial player of full moves.

\begin{ex}
  The category of elements of $\paraof{2}$ is the poset freely generated by the graph
  % \dam{bzzz: tu saurais faire un joli losange en baissant lt et en
  % montant rt?}  \tom{pas compris} \dam{c'est juste que a ligne du
  % haut du diagramme n'est pas très jolie, et on a un peu de mal à
  % retrouver le coup comme il est représenté en \ref{eq:para}, si tu
  % sais faire un losange pour la position d'arrivée, c'est mieux}
  % \dam{le commentaire n'était pas au bon endroit, ça ne devait pas
  % aider...}  \tom{ah, je comprends. je voyais pas trop comment faire
  % ça joliment. en fait il faudrait que $l$, $r$ et
  % $\id_{\paraof{2}}$ soient tout près les uns des autres, ce qui
  % rendrait le truc illisible.}
  \begin{center}
    \diag (1,.5) {%
      |(lt1)| l t d_1 = r t d_1 \& \& |(lt)| l t \& \& |(rt)| r t \& \& |(lt2)| l t d_2 = r t d_2 \\
      \& \& |(l)| l \& |(para)| \id_{\paraof{2}} \& |(r)| r \\ 
      |(ls1)| l s d_1 = r s d_1 \& \& \& |(ls)| l s = r s \&  \& \& |(ls2)| l s d_2 = r s d_2. %
    }{%
      (lt1) edge[identity] (ls1) %
      edge (lt) %
      edge[bend left=15] (rt) %
      (lt2) edge[identity] (ls2) %
      edge (rt) %
      edge[bend right=15,fore] (lt) %
      (ls1) edge (ls) %
      (ls2) edge (ls) %      
      (ls) edge (l) edge (r) %
      (lt) edge (l) %
      (rt) edge (r) %
      (l) edge (para) %
      (r) edge (para) %
    }
  \end{center}
  The two views corresponding to left and right half-forking are
  subcategories, and the object $\id_{\paraof{2}}$ `ties them
  together'.
\end{ex}

Presheaves on $\C_3$ are enough to
model full moves; to model closed-world moves, and in particular
synchronisation, we continue as follows.

Let $\G_4$ be $\G_3$, augmented with, for all $n$, $m$, $1 \leq i \leq
n$, and $1 \leq j \leq m$,
 \begin{itemize}
 \item a vertex $\taunimj$, and
 \item edges $\sender \colon \iotaposni \to \taunimj$ and $\receiver
   \colon \iotanegmj \to \taunimj$ ($\sender$ and $\receiver$
   respectively stand for `emission' and `reception').
 \end{itemize}
 \begin{defn}
   Let $\C_4$ be the category generated by $\G_4$, the previous
   relations, plus, for each $\iotaposni \xto{\sender} \taunimj
   \xot{\receiver} \iotanegmj$, the relation $\sender \rond s \rond
   d_i = \receiver \rond s \rond d_j$.
 \end{defn}
 This equation is the exact point where we enforce that a
 synchronisation involves an input and an output on the same channel,
 as announced in Example~\ref{ex:in}.

 \begin{ex}
   The category of elements of $\tauof{3}{3}{1}{1}$ is the preorder
   freely generated by the graph
  \begin{center}
    \diag(\diaghautici,.4){%
      |(et1)| \sender t  d_1     \& \&  \& \&  \\
      \& \& |(et)| \sender t \& \&  |(et3)| \sender t d_3 = %
      \receiver t d_1 \& \& \&  |(rt)| \receiver t \& \& |(rt2)| \receiver t d_2  \\
      \& |(et2)| \sender t  d_2 \&  \& \&  \\
      \\
      \& \& |(sender)| \sender  \&  \& \&  |(tau)| \id_{\tauof{3}{3}{2}{1}}  \& \& |(receiver)| \receiver \\
      \\
      |(es1)| \sender s  d_1     \& \&  \& \&  \\
      \& \& |(es)| \sender s  \& \&  %
      |(es3)| \sender s d_3 = \receiver s d_1 \& \& \& |(rs)|
      \receiver s \& \&
      |(rs2)| \receiver s d_2 \\
      \& |(es2)| \sender s d_2. \& \& \& %
    }{%
      (es2) edge (es) %
      (es1) edge (es) %
      (es3) edge (es) %
      (et2) edge (et) %
      (et1) edge (et) %
      (et3) edge (et) %
      (et) edge (sender) %
      (es) edge (sender) %
      (es1) edge[fore,identity] (et1) %
      (es2) edge[fore,identity] (et2) %
      (es3) edge[fore,identity] (et3) %
      (rs2) edge (rs) %
      (es3) edge (rs) %
      (rt2) edge (rt) %
      (et3) edge (rt) %
      (rt) edge (receiver) %
      (rs) edge (receiver) %
      (rs2) edge[fore,identity] (rt2) %
      (sender) edge[fore] (tau) %
      (receiver) edge[fore] (tau) %
    }%
  \end{center}
  Again, the two views corresponding to $\iotaof{3}{3}^+$ and
  $\iotaof{2}{1}^-$ are subcategories, and the new object
  $\tauof{3}{3}{2}{1}$ ties them together.
 \end{ex}

\subsection{Positions and moves}
We have now defined the base category $\C = \C_4$ on which the string
diagrams of Section~\ref{sec:string} are presheaves. More accurately
we have defined a sequence $\C_0 \into \ldots \into \C_4$ of
subcategories. 

\paragraph{Positions}
Positions are finite presheaves on $\C_1$, or
equivalently, finite presheaves on $\C_4$ empty except over
$\C_1$. 

\paragraph{Moves}
Basic moves should be essentially representable presheaves on objects
in $\ob (\C_2) \setminus \ob (\C_1)$. Recall however that basic moves
are defined as particular cospans in $\Chat$. This is also easy: in
the generating graph $\G_2$, each such object $c$ has exactly two
morphisms $s$ and $t$ into it, from objects, say, $[n_s]$ and $[n_t]$,
respectively. By Yoneda, these induce a cospan $[n_s] \xto{s} c
\xot{t} [n_t]$ in $\Chat$, which is the desired cospan. (Observe,
again, that only $\nun$ has $n_s \neq n_t$.)

Similarly, full moves either are basic moves, or are essentially
representable presheaves on objects in $\ob (\C_3) \setminus \ob
(\C_1)$, i.e., representables on some $\paran$. To define the expected
cospan, first observe that by the equation $l s = r s$, we obtain an
morphism $[n] \xto{s} \paraln \xto{l} \paran$, equal to $rs$, in $\Chat$.
This will form the first leg of the cospan. For the other, observe
that for each $n$ and $i \in n$, we obtain, by the equations $l t d_i
= l s d_i = r s d_i = r t d_i$ and by Yoneda, that the outermost part of
\begin{equation}
  \Diag (1,1) {%
    \pbk{lt}{t}{rt} %
  }{%
    |(star)| n \cdot \star \& |(rt)| [n]  \& \\
        |(lt)| [n] \& |(t)| n|n \& |(r)| \pararn \\
        \& |(l)| \paraln \& |(para)| \paran %
  }{%
    (star) edge[labelu={[d_i]_{i \in n}}] (rt) %
    edge[labell={[d_i]_{i \in n}}] (lt) %
    (rt) edge[labelar={t}] (r) %
    (lt) edge[labelbl={t}] (l) %
    (r) edge[labelr={r}] (para) %
    (l) edge[labeld={l}] (para) %
    (lt) edge (t) %
    (rt) edge (t) %
    (t) edge[dashed,labelar={\scriptstyle t}] (para) %
  }\label{eq:nparan}
\end{equation}
commutes in $\Chat$, where $n \cdot \star$ denotes an $n$-fold
coproduct of $\star$. Letting $n|n$ be the induced pushout, and the
dashed morphism $t$ be obtained by its universal property, we obtain the
desired cospan $[n] \xto{ls} \paran \xot{t} n|n.$

\begin{figure}[tp]
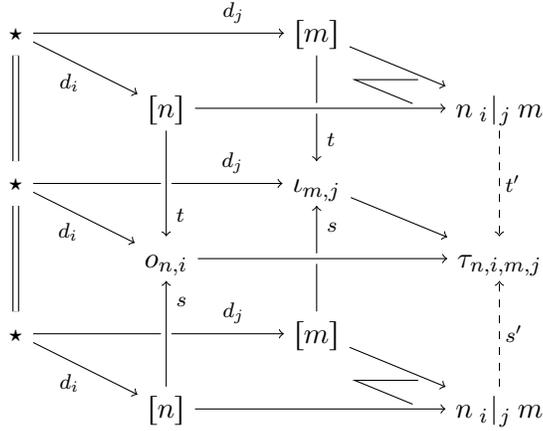

  \begin{center}
    \diagramme[diagorigins={1}{2}]{}{%
      \pbk{nii}{nmii}{mii} %
      \pbk{niv}{nmiv}{miv} %
    }{%
      |(starii)| \star \& \& |(mii)| [m] \\
      \& |(nii)| [n] \& \& |(nmii)| n \paraij m  \\
      |(g)| \star \&\& |(iotaneg)| \iotanegmj \\
      \& |(iotapos)| \iotaposni \& \& |(tau)| \taunimj \\
      |(stariv)| \star \& \& |(miv)| [m] \\
      \& |(niv)| [n] \& \& |(nmiv)| n \paraij m  %
    }{%
      % Ligne verticale 3
      (mii) edge node[right,pos=.8]{$\scriptstyle t$} (iotaneg) %
      (miv) edge node[right,pos=.8]{$\scriptstyle s$} (iotaneg) %
      % Carre 2
      (starii) edge[labelbl={d_i}]                         (nii) %
      (starii) edge node[above,pos=.8] {$\scriptstyle d_j$} (mii) %
      (nii) edge[fore] (nmii) %
      (mii) edge (nmii) %
      % Carre 4
      (stariv) edge[labelbl={d_i}]             (niv) %
      (stariv) edge node[above,pos=.8] {$\scriptstyle d_j$} (miv) %
      (niv) edge[fore] (nmiv) %
      (miv) edge (nmiv) %
      % Carre 3
      (g) edge[labelbl={d_i}]             (iotapos) %
      (g) edge node[above,pos=.8] {$\scriptstyle d_j$} (iotaneg) %
      (iotapos) edge[fore] (tau) %
      (iotaneg) edge (tau) %
      % Ligne verticale 1
      (starii) edge[identity] (g) %
      (stariv) edge[identity] (g) %
      % Ligne verticale 2
      (nii) edge[fore] node[right,pos=.8]{$\scriptstyle t$} (iotapos) %
      (niv) edge[fore] node[right,pos=.8]{$\scriptstyle s$} (iotapos) %
      % Ligne vertical 4
      (nmii) edge[dashed,labelr={t'}] (tau) %
      (nmiv) edge[dashed,labelr={s'}] (tau) %
    }
  \end{center}
  \caption{Construction of the synchronisation move}
\label{fig:sync}
\end{figure}

Finally, closed-world moves either are full moves, or are essentially
representable presheaves on some $\taunimj$. To define the expected
cospan, we proceed as in Figure~\ref{fig:sync}: compute the pushout
$n \paraij m$, and infer the dashed morphisms $s'$ and $t'$ to obtain the
desired cospan $n \paraij m \xto{s'} \taunimj \xot{t'} n \paraij m$.

\begin{rk}[Isomorphisms]\label{rk:iso}
  Moves are particular cospans in $\Chat$. For certain moves, the
  involved objects are representable, but not for others, like forking
  or synchronisation, whose final position is not representable. In
  the latter cases, our definition thus relies on a choice, e.g., of
  pushout in~\eqref{eq:nparan}.  Thus, let us be completely accurate:
  a move is a cospan which is isomorphic to one of the cospans chosen
  above, in $\Chat^{\cdot \ot \cdot \to \cdot}$, i.e., the category of
  functors from the category $\cdot \ot \cdot \to \cdot$ (generated by
  the graph with three objects and an edge from one to each of the
  other two) to $\Chat$.
\end{rk}

\subsection{Extended moves, plays, and relativisation}

The most delicate part of our formalisation of
Section~\ref{sec:diagrams} is perhaps the passage from moves to
extended moves. Recall from the paragraph above
Definition~\ref{def:play} that an extended move should be like a move
occurring in a larger position.

\paragraph{Moves with interfaces}
To formalise this idea, we first equip moves with interfaces, as
standard in graph rewriting~\cite{JLS}. Since moves are cospans, one
might expect that interfaces are cospans too. This may be done, but
there is a simpler, equivalent presentation. The route we follow here
might have to be generalised in order to handle more complex calculi
than CCS, but let us save the complications for later work.

Here, we define an \emph{interface} for a cospan $X \to M \ot Y$ to
consist of a presheaf $I$ and morphisms $X \ot I \to Y$ such that
\begin{equation}
  \diag{%
    I \& Y \\
    X \& M %
  }{%
    \sq{}{}{}{} %
  }\label{eq:cospanI}
\end{equation}
commutes, and $I$ has dimension 0, i.e., is empty except over $\C_0$,
i.e., consists only of channels. 

\begin{defn}
  A cospan equipped with an interface is called a \emph{cospan with
    interface}.
\end{defn}

Moves are particular cospans, and we now equip them with canonical
interfaces: all moves except channel creation preserve the set of channels,
the interface is then $n \cdot \star$, with the obvious inclusion. For
example, the less obvious case is $\paran$: we choose
\begin{center}
  \diag{%
    n \cdot \star \& {n|n} \\
    {[n]} \& \paran, %
  }{%
    \sq{}{}{}{} %
  }
\end{center}
where the upper map is as in~\eqref{eq:nparan}. For channel creation, we
naturally choose
\begin{center}
  \diag{%
    n \cdot \star \& {[n+1]} \\
    {[n]} \& \nun. %
  }{%
    \sq{[d_i]_{i \in n}}{}{}{} %
  }
\end{center}

\begin{defn}
  A \emph{move with interface} is one of these cospans with
  interface. The basic, full, or closed-world character is retained
  from the underlying move.
\end{defn}

\paragraph{Extended moves}
We now plug moves with interfaces into contexts, in the following
sense.
\begin{defn}
  A \emph{context} for a cospan with interface~\eqref{eq:cospanI} is a
  position $Z$, equipped with a morphism $I \to Z$.
% \dam{$I\into Z$ ? (ou virer les into sur le diagramme dessous ?)}
% \tom{heing? quels onto?}
% \dam{into, sorry: le diagramme juste en dessous a des into alors que
%   la def. non}
% \tom{ah. oui, mais c'est $I \into X$ qui est inj, pas $I \to
% Z$. oui?}
% \dam{oui, quel pibou je suis}
\end{defn}

From any cospan with interface $\mu$ as in~\eqref{eq:cospanI} and context $C
\colon I \to Z$, we construct the cospan $C[\mu]$ as in:
  \begin{center}
    \Diag(.5,1){%
      \pbk{X}{X'}{Z} %
      \pullback{Z}{M'}{M}{draw,-,fore} %
      \pullback[.8cm]{Z}{Y'}{Y}{draw,-,fore} %
    }{%
      \& |(Y)| Y \& \& |(Y')| {Y'} \\
      \& \ \& \\
      \& |(M)| M \& \& |(M')| M'  \\
      |(I)| I \&\& |(Z)| Z \\
      \& |(X)| X \& \& |(X')| X'. %
    }{%
      (Z) edge[into] (X') %
      edge (M') %
      edge (Y') %
      (I) edge[into] (X) %
      edge node[pos=.8,anchor=south] {$\scriptstyle C$} (Z) %
      edge (M) %
      edge (Y) %
      (Y) edge[fore] (Y') %
      (M) edge[fore] (M') %
      (X) edge (X') %
      (X') edge[dashed,into] (M') %
      (Y') edge[dashed,into] (M') %
      (X) edge[fore,into] (M) %
      (Y) edge[fore,into] (M) %
    }
  \end{center}
  I.e., we push the available morphisms out of $I$ along $C$, and
  infer the dashed morphisms, which form the desired cospan.

  \begin{defn}
    An \emph{extended move} is a cospan of the shape $C[\mu]$, for any
    move with interface $\mu$ and context $C$ as above.
  \end{defn}

  \begin{ex}
    Recall that $[2]$ is a position with one player knowing two channels. Recall from Figure~\ref{fig:sync} the pushout
    \begin{center}
      \Diag (1,1) {%
        \pbk{p1}{X}{p2} %
      }{%
        |(star)| \star \& |(p2)| {[2]} \\ %
        |(p1)| {[2]} \& |(X)| {2 \paradeuxun 2,} %
      }{%
        (star) edge[labelu={d_1}] (p2) %
        edge[labell={d_2}] (p1) %
        (p1) edge[labeld={p_1}] (X) %
        (p2) edge[labelr={p_2}] (X) %
      }
    \end{center}
    equivalently obtained as the pushout
    \begin{center}
      \Diag (1,1) {%
        \pbk{p1}{X}{p2} %
      }{%
        |(star)| \star + \star \& |(p2)| {\star + [2]} \\ %
        |(p1)| {[2]} \& |(X)| {2 \paradeuxun 2.} %
      }{%
        (star) edge[labelu={\id_\star + d_1}] (p2) %
        edge[labell={[d_1,d_2]}] (p1) %
        (p1) edge[labeld={p_1}] (X) %
        (p2) edge[labelr={[a_1,p_2]}] (X) %
      }
    \end{center}

    The base position of Figure~\ref{fig:exex} is thus $2 \paradeuxun
    2$. Recall also from~\eqref{eq:nparan} that $2|2$ denotes the
    position with two players both knowing two channels. Now, we have the
    forking move $[2] \into \paraof{2} \otni 2|2$. Equipping it with
    the interface
    $$[d_1,d_2] \colon \star + \star \to [2],$$
    and putting it in the context $\id_\star + d_1 \colon \star + \star \to
    \star + [2],$ (which happens to be the same as the interface), we obtain
      \begin{center}
    \Diag(.5,1){%
      \pbk{X}{X'}{Z} %
      \pullback{Z}{M'}{M}{draw,-,fore} %
      \pullback[.8cm]{Z}{Y'}{Y}{draw,-,fore} %
    }{%
      \& |(Y)| 2|2 \& \& |(Y')| {(2|2) \paradeuxun 2} \\
      \& \ \& \\
      \& |(M)| \paraof{2} \& \& |(M')| M  \\
      |(I)| \star + \star \&\& |(Z)| \star + [2] \\
      \& |(X)| [2] \& \& |(X')| 2 \paradeuxun 2. %
    }{%
      (Z) edge[into] (X') %
      edge (M') %
      edge (Y') %
      (I) edge[into,labelbl={[d_1,d_2]}] (X) %
      edge[into] node[pos=.8,anchor=south] {$\scriptstyle \id_\star + d_1$} (Z) %
      edge (M) %
      edge (Y) %
      (Y) edge[fore] (Y') %
      (M) edge[fore] (M') %
      (X) edge (X') %
      (X') edge[dashed,into] (M') %
      (Y') edge[dashed,into] (M') %
      (X) edge[fore,into] (M) %
      (Y) edge[fore,into] (M) %
    }
  \end{center}
  This formally constructs the first layer of
  Figure~\ref{fig:exex}. Constructing the whole play would be a little
  too verbose to be included here, but essentially straightforward.
  \end{ex}

\paragraph{Plays and relativisation}
We may now read Definition~\ref{def:play} again, this time in the
formal setting, to define plays. Similarly, the definition of
morphisms now makes rigorous sense, as well as
Proposition~\ref{prop:cat}.

\begin{myproof}[Proof of Proposition~\ref{prop:cat}] $\CE$ is the full subcategory of the arrow category of $\Chat$ whose
  objects are plays.\end{myproof}
%Tom: super bizarre, si je saute pas de ligne, ca fait n'imp...

Similarly, Section~\ref{subsec:relativisation} now makes rigorous
sense.

\section{Innocent strategies as sheaves}
Now that the category of plays is defined, we move on to defining
innocent strategies.  There is a notion of a Grothendieck
\emph{site}~\cite{MM}, which consists of a category equipped with a
(generalised) topology. On such sites, one may define a category of
sheaves, which are very roughly the presheaves that are determined
locally w.r.t.\ the generalised topology. We claim that there is a
topology on each $\CEX$, for which sheaves adequately model innocent
strategies.  Fortunately, in our setting, sheaves admit a simple
description, so that we can avoid the whole machinery. But sheaves
were the way we arrived at the main ideas presented here, because they
convey the right intuition: plays form a Grothendieck site, and the
states of innocent strategies should be determined locally.

In this section, we first define innocent strategies, and state the
spatial and temporal decomposition theorems. We then present our
coalgebraic interpretation of innocent strategies, i.e., we define a
polynomial endofunctor $\SynF$, and show that presheaves of finite
ordinals on views form a final $\SynF$-coalgebra. We then derive from this
a formal language and its interpretation in terms of innocent
strategies. We finally use this language to translate CCS with
recursive equations into innocent strategies.

\subsection{Innocent strategies}
\begin{defn}\label{defn:views}
  A \emph{view} is a finite, possibly empty `composition' $[n] \into
  V$ of (extended) basic moves in $\Cospan{\Chat}$, i.e., a play in
  which all the cospans are basic moves.
\end{defn}
When the composition is empty, we obtain $[n] \into [n]$, the
\emph{identity} view on $[n]$. We also note in passing that empty
presheaves cannot be views, i.e., $X \into \emptyset$ is never a view.

  \begin{ex}\label{ex:covpara}
    Forking~\eqref{eq:para} has two non-trivial views, namely the
    (left legs of) basic moves~\eqref{eq:paraviews}. Each of them embeds into forking:
    
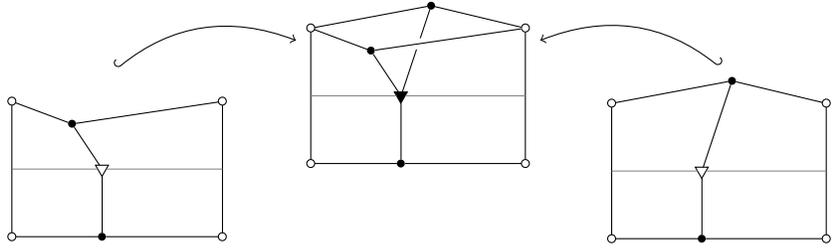
\begin{center}
    \begin{tikzpicture}
      \matrix(m)[stringdiag={.3}{.4}]{
        \& \& \&  \& \joueur{t_2} \\
        \canal{t0} \& \& \& \& \& \& \& \canal{t1}  \\ %\node[diagnode] (Y) {$Y$} ; \\
        \& \& \joueur{t_1}  \&  \\
        \& \ \& \\
        \coord{i0} \& \& \& \couppara{para} \& \& \& \& \coord{i1} \\ % \node[diagnode] (M) {$M$} ; \\
        \& \ \& \\
        \& \&  \\
        \canal{s0} \& \& \& \joueur{s} \& \& \& \& \canal{s1}  \\
      } ;
      \path[->] %
      (para) edge[-] (t_1) %
      (para) edge[-] (t_2) %
      (t0) edge[-] (t_1) %
      (t0) edge[-] (t_2) %
      (t1) edge[-,fore] (t_1) %
      (t1) edge[-] (t_2) %
      (s0) edge[-] (s) %
      (s1) edge[-] (s) %
      (s0) edge[-] (t0) %
      (s1) edge[-] (t1) %
      (s) edge[-] (para) %
      (i0) edge[-,gray,very thin] (para) %
      (i1) edge[-,gray,very thin] (para) %
      ; %
      % left half-forking
      \matrix(m1)[stringdiag={.3}{.4}] at (-4,-1) {
        \& \& \&   \\
    \canal{t0i} \& \& \& \& \& \& \& \canal{t1i} \\ %\node[diagnode] (Y) {$Y$} ; \\
    \& \& \joueur{t_1i}  \\
    \& \ \& \\
    \coord{i0i} \& \& \& \coupparacreux{parai} \& \& \& \& \coord{i1i} \\
    \& \ \& \\
    \& \&  \\
    \canal{s0i} \& \& \& \joueur{si} \& \& \& \& \canal{s1i} \\
  } ; %
  \path[->] %
    (t0i) edge[-] (t_1i) %
    (t1i) edge[-] (t_1i) %
    (s0i) edge[-] (si) %
    (s1i) edge[-] (si) %
    (s0i) edge[-] (t0i) %
    (s1i) edge[-] (t1i) %
    (si) edge[-] (parai) %
    (parai) edge[-] (t_1i) %
    (i0i) edge[-,gray,very thin] (parai) %
    (i1i) edge[-,gray,very thin] (parai) %
    ; %
      % right half-forking
    \matrix(m2)[stringdiag={.3}{.4}] at (4,-1) {%
    \& \& \&  \& \joueur{t_2ii} \\
    \canal{t0ii} \& \& \& \& \& \& \& \canal{t1ii}  \\ %\node[diagnode] (Y) {$Y$ii} ; \\
    \& \&   \&  \\
    \& \ \& \\
    \coord{i0ii} \& \& \& \coupparacreux{paraii} \& \& \& \& \coord{i1ii} \\ % \node[diagnode] (M) {$M$ii} ; \\
    \& \ \& \\
    \& \&  \\
    \canal{s0ii} \& \& \& \joueur{sii} \& \& \& \& \canal{s1ii} \\
  } ; %
  \path[->] %
    (t0ii) edge[-] (t_2ii) %
    (t1ii) edge[-] (t_2ii) %
    (s0ii) edge[-] (sii) %
    (s1ii) edge[-] (sii) %
    (s0ii) edge[-] (t0ii) %
    (s1ii) edge[-] (t1ii) %
    (sii) edge[-] (paraii) %
    (paraii) edge[-] (t_2ii) %
    (i0ii) edge[-,gray,very thin] (paraii) %
    (i1ii) edge[-,gray,very thin] (paraii) %
  ; %
  \path[->] (m1.north) edge[into,bend left] (m.160) %
  (m2.north) edge[linto,bend right] (m.20) %
  ; %
  \stringpoint{s1ii} %
  \end{tikzpicture}
    \end{center}
  \end{ex}
\begin{ex}
  In Figure~\ref{fig:exex}, the leftmost branch contains a view
  consisting of three basic moves: two $\paralof{2}$ and an output.
\end{ex}

\begin{defn}
  For any position $X$, let $\CVX$ be the full subcategory of $\CEX$
  consisting of views.
\end{defn}
More precisely, $\CVX$ consists of spans $U \otni Y \to X$ where $Y
\into U$ is a view.
\begin{defn}
  Let the category $\SS_X$ of \emph{innocent strategies} on $X$ be the
  category $\CVXhat$ of presheaves on $\CVX$.
\end{defn}
A possible interpretation is that for a presheaf $F \in \CVXhat$ and
view $V \in \CV_X$, $F (V)$ is the set of possible \emph{states} of
the strategy $F$ after playing $V$.

It might thus seem that we could content ourselves with defining only
views, as opposed to plays.  However, in order to define interactive
equivalences in Section~\ref{sec:inter}, we need to view innocent
strategies as (particular) presheaves on the whole of $\CEX$.

\begin{wrapfigure}{r}{0pt}
  \Diag(1.1,1){%
%    \twocell[20]{al}{E}{G}{}{cell=0,bend right} %
  }{%
    |(C)| \C \& \& \& |(D)| \D \\
    \& |(E)| \E \& %
  }{%
    (C) edge[labelu={F}] (D) %
    edge[bend right,labelbl={G}] node[pos=.4] (G) {} (E) %
    (D) edge[bend right] node[pos=.5] (al) {}
    node[pos=.2,anchor=north] {$\scriptstyle H$} (E) %
    edge[bend left,labelbr={K}] node[pos=.5] (br) {} (E) %
    (br) edge[cell={0},labelr={\scriptstyle \alpha'}] (al) %
    (al) edge[cell={0},bend right=10,labelu={\scriptstyle
      \varepsilon}] (G) (br) edge[cell={.1}] node[pos=.6,anchor=south]
    {$\scriptstyle \alpha$} (G) %
  }
\end{wrapfigure}
The connection is as follows. Recall from MacLane~\cite{MacLane:cwm}
the notion of \emph{right Kan extension}. Given functors $F$ and $G$
as on the right, a right Kan extension $\Ran_F (G)$ of $G$ along $F$
is a functor $H \colon \D \to \E$, equipped with a natural
transformation $\varepsilon \colon HF \to G$, such that for all functors
$K \colon \D \to \E$ and transformations $\alpha \colon KF \to G$,
there is a unique $\alpha' \colon K \to H$ such that $\alpha =
\varepsilon \vrond (\alpha' \rond \id_F)$, where $\vrond$ is
vertical composition of natural transformations.  Now, precomposition
with $F$ induces a functor $\Cat(F,\E) \colon \Cat (\D,\E) \to \Cat
(\C,\E)$, where $\Cat (\D,\E)$ is the category of functors $\D \to \E$
and natural transformations between them. When $\E$ is complete, right
Kan extensions always exist (and an explicit formula for our setting
is given below), and choosing one of them for each functor $\C \to \E$
induces a right adjoint to $\Cat (F,\E)$. Furthermore, it is known
that when $F$ is full and faithful, then $\varepsilon$ is a natural
isomorphism, i.e., $HF \iso G$.
\begin{prop}
  If $F$ is full and faithful, then $\Ran_F$ is a full essential
  embedding.
\end{prop}
\begin{myproof}
  First, let us show that $\Ran_F$ is essentially injective on
  objects. Indeed, assume $H = \Ran_F (G)$, $\Ran_F (G') = H'$, and $i
  \colon H \to H'$ is an isomorphism with inverse $k$.  We must
  construct an isomorphism $G \iso G'$.  Let $j \colon G \to G'$ be
  $\varepsilon_{G'} \vrond (i F) \vrond
  \inv{\varepsilon_G}$. Similarly, let $l \colon G' \to G$ be
  $\varepsilon_{G} \vrond (k F) \vrond \inv{\varepsilon_{G'}}$.  We
  have $$\begin{array}[t]{lcl}
    l \vrond j &=& \varepsilon_{G} \vrond (kF)  \vrond \inv{\varepsilon_{G'}} \vrond \varepsilon_{G'} \vrond (iF) \vrond \inv{\varepsilon_G} \\
    &=& \varepsilon_{G} \vrond (kF) \vrond (iF) \vrond \inv{\varepsilon_G} \\
    &=& \varepsilon_{G} \vrond ((k \vrond i) \rond F) \vrond \inv{\varepsilon_G} \\
    &=& \varepsilon_{G} \vrond \inv{\varepsilon_G} \\
    &=& \id_G.
  \end{array}$$
  Similarly, $j \vrond l = \id_{G'}$ and we have $G \iso G'$.

  To see that $\Ran_F$ is full, observe that for any $i \colon H \to
  H'$, with $H = \Ran_F (G)$ and $H' = \Ran_F (G')$, $j =
  \varepsilon_{G'} \vrond (i F) \vrond \inv{\varepsilon_G}$ is an antecedent
  of $i$ by $\Ran_F$.  Indeed, by definition, $\Ran_F (j)$ is the
  unique $i' \colon H \to H'$ such that $\varepsilon_{G'} \vrond (i' F) =
  j \vrond \varepsilon_G$. But the latter is equal to $\varepsilon_{G'}
  \vrond (i F)$, so $i' = i$.

  Finally, to show that $\Ran_F$ is faithful, consider $G, G' \colon
  \C \to \E$ and two natural transformations $i, j \colon G \to G'$
  such that $\Ran_F (i) = \Ran_F (j) = k$. Then, by construction of
  $k$, we have
  $$i \vrond \varepsilon_G = \varepsilon_{G'} \vrond (k F) = j \vrond \varepsilon_G.$$
  But, $\varepsilon_G$ being an isomorphism, this implies $i = j$ as
  desired.
\end{myproof}

Returning to views and plays, the embedding $i_X \colon \CVX \into
\CEX$ is full, so right Kan extension along $\op{i_X} \colon \op{\CVX}
\to \op{\CEX}$ induces a full essential embedding $\Ran_{\op{i_X}}
\colon \CVXhat \to \CEXhat$.  The (co)restriction of this essential
embedding to its essential image thus yields an essentially
surjective, fully faithful functor, i.e., an equivalence of
categories:
\begin{prop}\label{prop:synsem}
  The category $\SS_X$ is equivalent to the essential image of
  $\Ran_{\op{i_X}}$.
\end{prop}
The standard characterisation of right Kan extensions as
ends~\cite{MacLane:cwm} yields, for any $F \in \CVXhat$ and $U \in
\CEX$:
$$\Ran_{\op {i_X}}(F)(U) = \int_{V \in \CVX} F(V)^{\CEX(V,U)},$$
i.e., giving an element of $\Ran_{\op {i_X}}(F)$ on a play $U$ amounts
to giving, for each view $V$ and morphism $V \to U$, an element of
$F(V)$, satisfying some compatibility conditions. In
Example~\ref{ex:ran} below, we compute an example right Kan extension.

The interpretation of strategies in terms of states extends: for any
presheaf $F \in \CEXhat$ and play $U \in \CE_X$, $F (U)$ is the set of
possible \emph{states} of the strategy $F$ after playing $U$.  That
$F$ is in the image of $\Ran_{\op{i_X}}$ amounts to $F (U)$ being a
compatible tuple of states of $F$ after playing each view of $U$.

%  \begin{wrapfigure}{r}{0pt}
 %   \diag{\& |(x)| x \& \\ \& |(a)| a \& \\ |(y)| y \& \& |(z)| z}{(a) edge[-] (x) edge[-] (y) edge[-] (z)}
 % \end{wrapfigure}
\begin{ex}\label{ex:ran}
  Here is an example of a presheaf $F \in \CEXhat$ which is not
  innocent, i.e., not in the image of $\Ran_{\op{i_X}}$. Consider the
  position $X$ consisting of three players, say $x,y,z$, sharing a
  channel, say $a$. Let $X_x$ be the subposition with only $x$ and $a$,
  and similarly for $X_y$, $X_z$, $X_{x,y}$, and $X_{x,z}$.  Let $I_x
  = (\iotaneg{1,1} \otni X_x \into X)$ be the play where $x$ inputs on
  $a$, and similarly let $O_y$ and $O_z$ be the plays where $y$ and
  $z$ output on $a$, respectively. Let now $S_{x,y} =
  (\tauof{1}{1}{1}{1} \otni X_{x,y} \into X)$ be the play where $x$
  and $y$ synchronise on $a$ ($x$ inputs and $y$ outputs), and
  similarly let $S_{x,z}$ be the play where $x$ and $z$ synchronise on
  $a$. 

  Finally, we define a presheaf $F$ on $\E / X$ such that $F(S_{x,y})
  = 2$ is a two-element set, and $F (S_{x,z}) = \emptyset$. To define
  $F$ on other plays, the idea is to map any strict subplay of
  $S_{x,y}$ and $S_{x,z}$ to a one-element set $1$, and other plays to
  $\emptyset$. The cleanest technical way to do this seems to be as
  follows. The poset $\Exonze$ defined by
  \begin{center}
    \diag{%
      |(Oy)| O_y \& \& |(Ix)| I_x \& \& |(Oz)| O_z \\
      \& |(Sxy)| S_{x,y} \& \& |(Sxz)| S_{x,z} %
    }{%
      (Oy) edge (Sxy) %
      (Oz) edge (Sxz) %
      (Ix) edge (Sxy) edge[fore] (Sxz) %
    }  
  \end{center}
  fully embeds into $\E / X$, via, say $i_{11}$. Let $F_0$ be the
  presheaf on $\Exonze$ defined by:
  \begin{center}
    \Diag{%
      \path[<-,draw] %
      (Oy) edge (Sxy) %
      (Oz) edge (Sxz) %
      (Ix) edge (Sxy) edge[fore] (Sxz) %
      ; %
    }{%
      |(Oy)| 1 \& \& |(Ix)| 1 \& \& |(Oz)| 1 \\
      \& |(Sxy)| 2 \& \& |(Sxz)| \emptyset{.\!\!} %
    }{%
    }  
  \end{center}
  We now let $F = \Ran_{\op{i_{11}}} (F_0)$. Because $i_{11}$ is fully
  faithful, $F$ coincides with $F_0$ on the plays of $\Exonze$, as
  desired. 

  Now, $F$ fails to be innocent on two counts.  First, since $x$ and
  $y$ accept to input and output in only one way, it is non-innocent
  to accept that they synchronise in more than one way. Formally,
  $S_{x,y}$ has two non-trivial views, $I_x$ and $O_y$, so since $F$
  maps identity views to a singleton, $F (S_{x,y})$ should be
  isomorphic to $F (I_x) \times F (O_y) = 1 \times 1 = 1$. The second
  reason why $F$ is not innocent is that, since $x$ and $z$ accept to
  input and output, $F$ should accept that they synchronise. Formally,
  $F (S_{x,z})$ should also be a singleton. This altogether models the
  fact that in CCS, processes do not get to know with which other
  processes they synchronise.

  The restriction of $F$ to $\CVX$, i.e., $F' = F \rond \op{i_X}$, in
  turn has a right Kan extension $F''$, which is innocent. (In
  passing, the unit of the adjunction $\Cat (\op{i_X},\Set) \dashv
  \Ran_{\op{i_X}}$ is a natural transformation $F \to F''$.) To
  conclude this example, let us compute $F''$. First, $F'$ only
  retains from $F$ its values on views. So, if $X_x$ denotes the empty
  view on $X_x$, $F' (X_x) = 1$, and similarly $F' (X_y) = F' (X_z) =
  1$. Furthermore, $F' (I_x) = F' (O_y) = F' (O_z) = 1$. Finally, for
  any view $V$ not isomorphic to any of the previous ones, $F' (V) =
  \emptyset$. So, recall that $F''$ maps any play $U \otni Y \into X$
  to $\int_{V \in \CVX} F' (V)^{\CEX(V,U)}$.  So, e.g., since the
  views of $S_{x,y}$ are subviews of $I_x$ and $O_y$, we have $F''
  (S_{x,y}) = F' (I_x) \times F' (O_y) = 1$. Similarly, $F'' (S_{x,z})
  = 1$.  But also, for any play $U$ such that all views $V \to U$ are
  subviews of either of $I_x$, $O_y$, or $O_z$, we have $F'' (U) =
  1$. Finally, for any play $U$ such that there exists a view $V \to
  U$ which is not a subview of any of $I_x$, $O_y$, or $O_z$, we have
  $F'' (U) = \emptyset$.  
\end{ex}
One way to understand Proposition~\ref{prop:synsem} is to view
$\CVXhat$ as the syntax for innocent strategies: presheaves on views
are (almost) infinite terms in a certain syntax (see
Section~\ref{subsec:lang} below).
On the other hand, seeing them as presheaves on plays will allow us to
consider their global behaviour: see Section~\ref{sec:inter} when we
restrict to the closed-world game.
Thus, right Kan extension followed by restriction to closed-world will
associate a semantics to innocent strategies.

\begin{rk}
  The relevant Grothendieck topology on $\CEX$ says, roughly, that a
  play is covered by its views. Any sheaf for this topology is
  determined by its restriction to $\CVX$, for its elements on any
  non-view play $U$ are precisely amalgamations of its elements on
  views of $U$. Right Kan extension just computes these amalgamations
  in the particular case of a topology derived from a full
  subcategory, here views.
\end{rk}

So, we have defined for each $X$ the category $\SS_X$ of innocent
strategies on $X$. This assignment is actually functorial $\op\B \to
\CAT$, as follows (where $\CAT$ is the large category of locally small
categories). Any morphism $f \colon Y \to X$ induces a functor $f_!
\colon \CVY \to \CVX$ mapping $(V \otni Z \to Y)$ to $(V \otni Z \to Y
\to X)$. Precomposition with $\op{(f_!)}$ thus induces a functor
$\SS_f \colon \CVXhat \to \CVYhat$.
\begin{prop}
  This defines a functor $\SS \colon \op\B \to \CAT$.
\end{prop}
\begin{myproof}
  A straightforward verification.
\end{myproof}

But there is more: for any position, giving a strategy for each player
in it easily yields a strategy on the whole position. We call this
\emph{amalgamation} of innocent strategies (because the functor $\SS$
is indeed a \emph{stack}~\cite{Vistoli}, and this is a particular case
of amalgamation in that stack). Formally, consider any subpositions
$X_1$ and $X_2$ of a given position $X$, inducing a partition of the
players of $X$, i.e., such that $X_1 \cup X_2$ contains all players of
$X$, and $X_1 \cap X_2$ contains none.  Then $\CVX$ is isomorphic to
the coproduct $\CV_{X_1} + \CV_{X_2}$. (Indeed, a view contains in
particular an initial player in $X$, which forces it to belong either
in $\CV_{X_1}$ or in $\CV_{X_2}$.)
\begin{defn}
  Given innocent strategies $F_1$ on $X_1$ and $F_2$ on $X_2$, let
  their \emph{amalgamation} be their copairing $$[F_1,F_2] : \op{\CVX}
  \iso (\CV_{X_1} + \CV_{X_2})^{op} \iso \op{\CV_{X_1}} +
  \op{\CV_{X_2}} \to \Set.$$
\end{defn}
By universal property of coproduct:
\begin{prop}\label{prop:spatial}
  Amalgamation yields an isomorphism of categories $$\CVXhat \iso
  \widehat{\CV_{X_1}} \times \widehat{\CV_{X_2}}.$$
\end{prop}

\begin{ex}
  Consider again the position $X$ from Example~\ref{ex:ran}, and let
  $X_{y,z}$ be the subposition with only $y$ and $z$.  We have $\CVX
  \equi (\CV_{X_x} + \CV_{X_{y,z}})$, which we may explain by hand as
  follows.  A view on $X$ has a base player, $x$, $y$, or $z$, and so
  belongs either in $\CV_{X_x}$ or in $\CV_{X_{y,z}}$. Furthermore, if
  $V$ is a view on $x$ and $W$ is a view on $y$, then $\CVX (V,W) =
  \emptyset$ (and similarly for any pair of distinct players in $X$).

  Now, recall $F'$, the restriction of $F$ to $\CVX$. We may define
  $F_x \colon \op{\CV_{X_x}} \to \Set$ to be the restriction of $F'$
  along the (opposite of the) embedding $\CV_{X_x} \into \CVX$, and
  similarly $F_{y,z}$ to be the restriction of $F'$ along
  $\CV_{X_{y,z}} \into \CVX$. We have obviously $F' = [F_x,
  F_{y,z}]$.
\end{ex}

Analogous reasoning leads to what we call spatial decomposition. For
any $X$, let $\Pl (X) = \sum_n X([n]),$ i.e., the set of pairs
$(n,x)$, where $x$ is a player in $X$, knowing $n$ channels.  
\begin{thm}\label{thm:spatial}
  We have $\CVXhat \iso \prod_{(n,x) \in \Pl (X)} \CVnhat$.
\end{thm}

Again, this is a particular case of amalgamation in the stack $\SS$,
but we do not need to spell out the definition here.

\subsection{Temporal decomposition}
Let us now describe \emph{temporal} decomposition.  Recall that basic
moves are left and right half-forking~\eqref{eq:paraviews}, input,
output, tick, and channel creation.
  \begin{defn}
    Let $\MMM$ be the graph with vertices all natural numbers $n$, and
    with edges $n \to n'$ all (isomorphism classes of) basic moves $M
    \colon [n] \to [n']$.
  \end{defn}
  Recall from Remark~\ref{rk:iso} that the notion of isomorphism considered
  here is that of an isomorphism of cospans in $\Chat$.

%   For any graph $G$ and vertex $v \in G$, we let $v / G$ be the graph
%   with vertices  $p \colon v \to v'$ from $v$ in $G$, and with
%   edges from $p$ to $q \colon v \to v''$ all edges $e \colon v' \to
%   v''$ of $G$ such that $q$ is $p$, followed by $e$.

\begin{defn}
  Let $\MMM_n$ be the set of edges from $n$ in $\MMM$.
\end{defn}

For stating the temporal decomposition theorem, we need a
standard~\cite{Jacobs} categorical construction, the category of
families on a given category $\C$. First, given a set $X$, consider
the category $\Fam (X)$ with as objects $X$-indexed families of sets
$Y = (Y_x)_{x \in X}$, and as morphisms $Y \to Z$ families $(f_x
\colon Y_x \to Z_x)_{x \in X}$ of maps. This category is equivalently
described as the slice category $\Set / X$. To see the correspondence,
consider any family $(Y_x)_{x \in X}$, and map it to the projection
function $\sum_{x \in X} Y_x \to X$ sending $(x, y)$ to $x$.
Conversely, given $f \colon Y \to X$, let, for any $x \in X$, $Y_x$ be
the fibre of $f$ over $x$, i.e., $\inv{f} (x)$.

Generalising from sets $X$ to small categories $\C$, $\Fam (\C)$ has
as objects families $p \colon Y \to \ob(\C)$ indexed by the objects of
$\C$.  Morphisms $(Y, p) \to (Z, q)$ are pairs of $u \colon Y \to Z$
and $v \colon Y \to \mor(\C)$, where $\mor (\C)$ is the set of
morphisms of $\C$, such that $\dom \rond v = p$, and $\cod \rond v = q
\rond u.$ Thus, any element $y \in Y$ over $C \in \C$ is mapped to
some $u (y) \in Z$ over $C' \in \C$, and this mapping is labelled by
a morphism $v (y) \colon C \to C'$ in $\C$. The obtained category is
locally small.

Further generalising, for $\C$ a locally small category, we may define
$\Fam (\C)$ in exactly the same way (with $Y$ still a set), and the
obtained category remains locally small.

The temporal decomposition theorem is:
  \begin{thm}\label{thm:temp}
    There is an equivalence of categories $$\SS_n \equi \Fam
    \left(\prod_{M \in \MMM_n}\SS_{\cod (M)}\right ).$$
 \end{thm}
 The main intuition is that an innocent strategy is determined up to
 isomorphism by (i) its initial states, and (ii) what remains of them
 after each possible basic move.
%
%  \dam{tu ne voudrais pas dire `full' move? (en tout cas dans
%    l'intro, tu disais que full-move sert à avoir la temporal
%    decomposition)} %
%  \tom{oui, j'ai changé parce que c'est pareil mais moins verbeux avec
%    les coups basiques. Il faut peut-être modifier l'intro. Je me
%    demande à quel point ça vaut le coup de virer carrément la notion
%    de coup plein en mettant le forking au même niveau que la
%    synchro. Ça m'a l'air raté conceptuellement, mais je vois pas
%    encore ce que ça casse. Y a des preuves qui seraient moins faciles
%    et on en chierait un peu plus pour définir un critère de
%    correction. Bon je préfère pas le faire pour l'instant.}%
%  \dam{en effet, mais ce n'est plus le moment...}
% %
 The family construction is what permits innocent strategies with
 several possible states over the identity play.

 \begin{myproof}[Proof sketch]
 For general reasons, we have:
$$
\begin{array}[t]{lcl}
  \Fam
  \left(\prod_{M \in \MMM_n}\SS_{\cod (M)}\right ) 
  & =  &
  \Fam \left (\prod_{M \in \MMM_n}[\op{\CV_{\cod (M)}},\Set] \right ) \\
  & \iso  &
  \Fam \left (\left[\op{\sum_{M \in \MMM_n}{\CV_{\cod (M)}}},\Set\right] \right ) \\
    & \equi &    
    \left[\op{\sum_{M \in \MMM_n} \CV_{\cod (M)}},\Set\right] \downarrow
    \Delta,
  \end{array}
$$
where $\Delta \colon \Set \to [\op{\sum_{M \in \MMM_n} \CV_{\cod
    (M)}},\Set]$ maps any set $X$ to the constant presheaf mapping any
object to $X$ and any morphism to the identity.

By definition, the last category is a lax pullback
  \begin{center}
    \Diag (1,1) {%
      \path[<-,draw] %
      (Vns) edge[identity] (Vnsi) %
      edge[shorten <=.1cm,labell={\Delta}] (un) %
      (un) edge[labeld={}] (Vn) %
      (Vnsi) edge[shorten <=.1cm,labelr={}] (Vn) %
      ; %
      \twocell[.5]{Vnsi}{Vn}{un}{}{cell=0,bend
        right,labelal={}} %
    }{%
      |(Vns)| \left[\op{\sum_{M \in \MMM_n} \CV_{\cod (M)}},\Set\right] \&  |(Vnsi)|  \left[\op{\sum_{M \in \MMM_n} \CV_{\cod (M)}},\Set\right] \\
      |(un)| \Set \& |(Vn)| { \left[\op{\sum_{M \in \MMM_n} \CV_{\cod
            (M)}},\Set\right] \downarrow \Delta} %
    }{%
    }
  \end{center}
  in $\CAT$. 

  Now, any basic move $M \colon n \to n'$ induces a functor $(- \rond
  M) \colon \CVni \to \CVn$, mapping any view $V \in \CVni$ to $V
  \rond M$ (with composition in $\Cospan{\Chat}$).  We show that the
  square
  \begin{equation}
    \Diag (1,1) {%
      \twocell[.5]{Vnsi}{Vn}{un}{}{cell=0,bend
        right,labelal={\lambda}} %
    }{%
      |(Vns)| {\sum_{M \in \MMM_n} \op\CV_{\cod (M)}} \&  |(Vnsi)| {\sum_{M \in \MMM_n} \op\CV_{\cod (M)}} \\
      |(un)| 1 \& |(Vn)| \op\CVn %
    }{%
      (Vns) edge[identity] (Vnsi) %
      edge[shorten <=.1cm,labell={!}] (un) %
      (un) edge[labeld={\name{\id_{[n]}}}] (Vn) %
      (Vnsi) edge[shorten <=.1cm,labelr={[- \rond M]_{M \in
          \MMM_n}}] (Vn) %
    }\label{eq:laxpsht}
  \end{equation}
  is a lax pushout in $\Cat$, where $\lambda_{M, V} \colon \id_{[n]}
  \to M \rond V$, seen in $\CVn$, is the obvious inclusion, which for
  general reasons is mapped by the hom-2-functor $\CAT (-,\Set)$ to a
  lax pullback. But $\CAT (!,\Set) = \Delta$ and $\CAT (\id,\Set) =
  \id$, so we obtain a canonical isomorphism of lax pullbacks $$\SSn =
  [\op\CVn,\Set] \iso \left[\op{\sum_{M \in \MMM_n} \CV_{\cod
      (M)}},\Set\right] \downarrow \Delta.$$ More detail is in
  Appendix~\ref{sec:proofEE}.
\end{myproof}

 \begin{rk}
   The theorem almost makes innocent strategies into a \emph{sketch}
   (on the category with positions as objects, finite compositions of
   extended moves as morphisms, and the $\MMM_X$'s as distinguished
   cones). Briefly, being a sketch would require a bijection of sets
   $\SS_n \iso \prod_{M \in \MMM_n}\SS_{\cod (M)}.$ Here, the
   bijection becomes an equivalence of categories, and the family
   construction sneaks in.
 \end{rk}

 \subsection{Innocent strategies as a terminal coalgebra}
Temporal decomposition gives
$$\SS_n \equi \Fam \left( \prod_{M \in \MMM_n} \SS_{\cod (M)}\right),$$
for all $n$. Considering a variant of this formula as a system of
equations will lead to our interpretation of CCS.  The first step is
to replace $\Set$ with $\FinOrd$, the category of finite ordinals and
monotone functions. The proof applies \emph{mutatis mutandis} and we
obtain an equivalence, which, because both categories are skeletal, is
an isomorphism:
\begin{equation}
\label{eq:syntax}
  \CVnhatf \iso \Fam_f \left( \prod_{M \in \MMM_n} \OPsh{\CV_{\cod (M)}}\right),  
\end{equation}
where
\begin{itemize}
\item $\Fam_f$ is the same as $\Fam$ but with finite families, i.e.,
  for any category $\C$, $\ob(\Fam_f (\C)) = \sum_{I \in \FinOrd} (\ob
  (\C))^I = (\ob (\C))^*$ is the set of finite words over objects of
  $\C$, also known as the free monoid on $\ob (\C)$;
\item and for any category $\C$, $\OPsh \C$ denotes the functor
  category $[\op\C, \FinOrd]$.
\end{itemize}

\begin{rk}
  Recall that in the proof of Theorem~\ref{thm:temp}, $\Fam$ arises
  from the `constant presheaf' functor $\Delta \colon \Set \to
  \hat{-}$, with $-$ a complicated category. This functor itself is
  equal to restriction along $- \to 1$, via $\hat{1} \iso
  \Set$. Replacing $\Set$ with $\FinOrd$ thus replaces $\Delta$ with
  the analogous functor $\FinOrd \to \OPsh {-}$, via $\OPsh {1} \iso
  \FinOrd$, and thus $\Fam$ with $\Fam_f$.

  Furthermore, because $\FinOrd$ embeds into $\Set$, the special
  strategies of $\CVnhatf$ embed into $\SSn$.
\end{rk}

Then, taking advantage of the fact that $\FinOrd$ is a small category,
we consider its set $\FinOrd_0$ of objects, i.e., finite ordinals, and
the endofunctor $\SynF$ on $\Set / \FinOrd_0$ defined on any family of
sets $X = (X_i)_{i \in \FinOrd_0}$ by:
$$(\SynF (X))_n = \sum_{I \in \FinOrd_0} \left( \prod_{M \in \MMM_n} X_{\cod (M)}\right)^I,$$
where we abusively confuse $[n'] = \cod (M)$ and the natural number
$n'$ itself. The isomorphism~\eqref{eq:syntax} becomes $$\ob
(\CVnhatf) \iso (\SynF (\ob (\OPsh{\CV_{-}})))_{n}.$$

We may decompose $\SynF$ as follows.  Consider the endofunctor on
$\Set / \linebreak \FinOrd_0$ defined by $(\deriv X)_n = \prod_{M \in
  \MMM_n} X_{\cod (M)}$, for any family $X$. We obviously have:
\begin{lem}
  $\SynF$ is equal to the composite $(\deriv -)^*$.
\end{lem}

This endofunctor is polynomial~\cite{Kock01012011} and we now give a
characterisation of its final coalgebra.  The rest of this subsection
is devoted to proving:
\begin{thm}\label{thm:coalg}
  The family $\ob(\OPsh {\CV_n})$ formed for each $n$ by (the objects of)
  $\OPsh {\CV_n}$ is a terminal coalgebra for $\SynF$.
\end{thm}

Consider any $\SynF$-coalgebra $a \colon X \to \SynF X$.

We define by induction on $N$ a sequence of maps $f_N \colon X \to
\CVhatf$, such that for any view $V$ of length less than $N$ (i.e.,
with less than $N$ basic moves), and any $N' > N$, $f_{N'} (x) (V) =
f_{N} (x) (V)$, and similarly the action of $f_{N} (x)$ on morphisms
is the same as that of $f_{N'} (x)$.

To start the induction, take $f_0 (x)$ to be the strategy mapping
$\id_{[n]}$ to $\pi (a (x))$, i.e., the length of $a (x) \in \sum_{I
  \in \FinOrd_0} ((\deriv X)_n)^I$, and all other views to $0$.

Furthermore, given $f_N$, define $f_{N+1}$ to be
$$X \xto{a} \SynF X \xto{\SynF (f_N)} \SynF(\CVhatf) \xto{\iso} \CVhatf,$$
where the equivalence is by temporal decomposition.

Unfolding the definitions yields:
\begin{lem}\label{lem:unfold}
  Consider any $x \in X_n$, and $a (x) = (z_1, \ldots, z_k)$. For any
  move $M \colon n \to n'$ and view $V \colon n' \to n''$ of length at
  most $N$, and for any $i \in k$, $f_{N+1} (x) (V \rond M) = \sum_{i
    \in k} f_N (z_i (M)) (V)$.
\end{lem}

For any $x \in X_n$, we have a sequence $f_0 (x) \into f_1 (x) \into
\ldots f_N (x) \into f_{N+1} (x) \into \ldots$ which is pointwise
stationary. This sequence thus has a colimit in $\CVnhatf$, the
presheaf mapping any view $V$ of length $N$ to $f_N (V)$ (or
equivalently $f_{N'} (V)$ for any $N' \geq N$), which allows us to
define:
\begin{defn}
  Let $f \colon X \to \CVhatf$ map any $x \in X_n$ to $\bigcup_{N} f_N
  (x)$.
\end{defn}

By construction, we have
\begin{lem}
  The following diagram commutes:
  \begin{center}
    \diag{%
      |(X)| X \& |(FX)| \SynF X \\
      |(CVhatf)| \CVhatf \& |(FCVhatf)| \SynF (\CVhatf). %
    }{%
      (X) edge[labelu={a}] (FX) %
      edge[labell={f}] (CVhatf) %
      (FX) edge[labelr={\SynF(f)}] (FCVhatf) %
      (FCVhatf) edge[labelu={\iso}] (CVhatf) %
    }
  \end{center}
\end{lem}

\begin{lem}
  The map $f$ is a morphism of $\SynF$-coalgebras.
\end{lem}
\begin{myproof}
  Let, for any innocent strategy $S \in \CVnhatf$ and $i \in S
  (\id_{[n]})$, $\restr{S}{i}$ be the strategy mapping any view $V$ to
  the fibre over $i$ of $S (V) \to S (\id_{[n]})$. Using the notations
  of Lemma~\ref{lem:unfold}, we must show that for any $i \in k$, we
  have $\restr{(f (x))}{i} (V \rond M) = f (z_i (M)) (V)$.  But
  Lemma~\ref{lem:unfold} entails that $f (x) (V \rond M) \to f (x)
  (\id_{[n]})$ is actually the coproduct over $i' \in k$ of all $f
  (z_{i'} (M)) (V) \xto{!} 1 \xto{i'} \pi(a (x))$, so its fibre over
  $i$ is indeed $f (z_{i} (M)) (V)$.
\end{myproof}

\begin{lem}
  The map $f$ is the unique map $X \to \CVhatf$ of $\SynF$-coalgebras.
\end{lem}
\begin{myproof}
  Consider any such map $g$ of coalgebras. It must be such that $g (x)
  (\id_{[n]}) = \pi (a (x))$, and furthermore, using the same notation
  as before, for any $i \in k$ $\restr{(g (x))}{i} (V \rond M) = g
  (z_i (M)) (V)$, which imposes by induction that $f = g$.
\end{myproof}

The last two lemmas directly entail Theorem~\ref{thm:coalg}.

\subsection{Languages}\label{subsec:lang}

A consequence of Theorem~\ref{thm:coalg} is that the family
$\OPsh{\CV_n}$ supports the operations of the grammar
\begin{mathpar}
\inferrule{\ldots \  n \vdash F_i \  \ldots \ (\forall i \in I)}{
n \vdash \sum_{i \in I} F_i}~(I \in \FinOrd_0) \and
\inferrule{\ldots \ n' \vdash F_{M} \ \ldots \ {(\forall M \colon [n] \to [n'] \in \MMM)}%
}{
n \vdash \langle M \mapsto F_{M}\rangle 
}~\cdot
\end{mathpar}
Here, $n \vdash F$ denotes a presheaf of finite ordinals on $\CV_n$.
The interpretation is as follows: given presheaves $F_1, \ldots,
F_{I}$, for $I \in \FinOrd_0$, the first rule constructs the finite
coproduct $\sum_{i \in I} F_i$ of presheaves (finite coproducts exist in
$\OPsh{\CV_n}$ because they do in $\FinOrd$). In particular, when $I$
is the empty ordinal, we sum over an empty set, so the rule
degenerates to
\begin{mathpar}
  \inferrule{ }{n \vdash \emptyset}~\cdot
\end{mathpar}
In terms of presheaves, this is just the constantly empty presheaf.

For the second rule, if for all basic $M \colon [n] \to [n']$, we are given $F_{M}
\in \OPsh{\CVni}$, then $\langle M \mapsto
F_{M}\rangle$ denotes the image under~\eqref{eq:syntax} of
$$(1, 1 \mapsto M \mapsto F_{M}).$$
Here, we provide an element of the right-hand side
of~\eqref{eq:syntax}, consisting of the finite ordinal $I = 1 =
\ens{1}$, and the function mapping $M$ to $F_{M} \in \OPsh{\CVni}$ (up
to currying). That was for parsing; the intuition is that we construct
a presheaf with one initial state, $1$, which maps any view starting
with $M$, say $V \rond M$, to $F_{M} (V)$. Thus the $F_{M}$'s specify
what remains of our presheaf after each possible basic move. In
particular, when all the $F_{M}$'s are empty, we obtain a
presheaf which has an initial state, but which does nothing beyond
it. We abbreviate it as $0 = \langle \_ \mapsto \emptyset \rangle$.

\subsection{Translating CCS}
It is rather easy to translate CCS into this language.  First, define
CCS syntax by the natural deduction rules in Figure~\ref{fig:ccs},
where $\Names$ and $\Vars$ are two fixed, disjoint, and infinite sets
of \emph{names} and \emph{variables}; $\Xi$ ranges over finite
sequences of pairs $(x \colon n)$ of a variable $x$ and its
\emph{arity} $n \in \FinOrd_0$, such that the variables are pairwise
distinct; $\Gam$ ranges over finite sequences of pairwise distinct
names; there are two judgements: $\Gam \vdash P$ for \emph{global}
processes, $\Xi;\Gam \vdash P$ for \emph{open} processes.  Rule
\textsc{Global} is the only rule for forming global processes, and
there $\Xi = (x_1 \colon \card{\Del_1}, \ldots, x_n \colon
\card{\Del_n})$. Finally, $\alpha$ denotes $a$ or $\abar$, for $a \in
\Names$, and $\floor{a} = \floor{\abar} = a$.
\begin{figure}[t]
  \centering
  \begin{mathpar}
\inferrule[\textsc{CCSApp}]{ }{\Xi ; \Gam \vdash x (a_1, \ldots, a_{n})}~((x \colon n) \in \Xi \mbox{\ and\ } a_1, \ldots, a_{n} \in \Gam)
  \and
\inferrule{\Xi; \Gam, a \vdash P 
}{ %
  \Xi;\Gam \vdash \nu a.P %
}~\mbox{($a \notin \Gam$)}
\and
\inferrule{\Xi; \Gam \vdash P \\
\Xi; \Gam \vdash Q 
}{ %
  \Xi;\Gam \vdash P|Q %
}
\and
\inferrule{%
  \ldots \\ \Xi; \Gam \vdash P_i  \\ \ldots \\ (\forall i \in I)
}{ %
  \Xi;\Gam \vdash \sum_{i \in I} \alpha_i.P_i %
}~(I \in \FinOrd_0 \mbox{\ and\ } \forall i \in I, \floor{\alpha_i} \in \Gam)
\and
\inferrule[\textsc{Global}]{\Xi ; \Del_1 \vdash P_1 \\ \ldots \\ \Xi ; 
  \Del_n \vdash P_n \\
  \Xi ; \Gam \vdash P 
}{
  \Gam \vdash \recin{x_1 (\Del_1) \eq P_1, \ldots,
  x_n (\Del_n) \eq P_n}{P} %
}
\end{mathpar}
  \caption{CCS syntax}
  \label{fig:ccs}
\end{figure}

First, we define the following (approximation of a) translation on
open processes, mapping each open process $\Xi; \Gam \vdash P$ to
$\transl{P} \in \OPsh{\CV_n}$, for $n = \card{\Gam}$.  This
translation ignores the recursive definitions, and we will refine it
below to take them into account. We proceed by induction on $P$,
leaving contexts $\Xi; \Gam$ implicit:
 $$\begin{array}[t]{r@{\ \mapsto\ }lll}
 x (a_1, \ldots, a_{k}) & \emptyset \\
% 0 & \langle \emptyset \rangle &  \\
   P|Q & \langle \begin{array}[t]{r@{\ \mapsto\ }l}
    \paraln & \transl{P}, \\
    \pararn & \transl{Q}, \\
    \_ & \emptyset \ \ \rangle 
  \end{array}     
  \end{array}
  \begin{array}[t]{r@{\ \mapsto\ }lll}
  \nu a.P &  \langle \nuof{n} \mapsto \transl{P}, \_ \mapsto \emptyset \rangle \\
  \sum_{i \in I} \alpha_i.P_i &  \langle \!
  \begin{array}[t]{r@{\ \mapsto\ }l}
    (\iotaposnj &  \sum_{k \in I_{\overline{j}}} \transl{P_k}, \\
    \iotanegnj &  \sum_{k \in I_{j}} \transl{P_k}\ )_{j \in n}, \\
    \_ &  \emptyset\ \ 
    \rangle.
  \end{array}\end{array}$$
Let us explain intuitions and notation.  In the first case, we assume
implicitly that $(x \colon k) \in \Xi$; the intuition is just that we
approximate variables with empty strategies.  Next, $P|Q$ is
translated to the strategy with one initial state, which only accepts
left and right half-forking first, and then lets its avatars play
$\transl{P}$ and $\transl{Q}$, respectively. Similarly, $\nu a.P$ is
translated to the strategy with one initial state, accepting only the
channel creation move, and then playing $\transl{P}$.  In the last
case, the guarded sum $\sum_{i \in I} \alpha_i.P_i$ is translated to
the strategy with one initial state, which
\begin{itemize}
\item accepts input on any channel $a$ when $\alpha_i = a$ for some $i
  \in I$, and output on any channel $a$ when $\alpha_i = \abar$ for
  some $i \in I$;
\item after an input on $a$, plays the sum of all $\transl{P_i}$'s
  such that $\alpha_i = a$; and after an output on $a$, plays the sum
  of all $\transl{P_i}$'s such that $\alpha_i = \abar$.
\end{itemize}
Formally, in the definition, we let, for all $j \in n$,
$I_{\overline{j}} = \ens{i \in I \aalt \alpha_i = \overline{a_j}}$ and
$I_j = \ens{i \in I \aalt \alpha_i = a_j}$.  In particular, if $I =
\emptyset$, we obtain~$0$.

Thus, almost all translations of open processes have exactly one
initial state, i.e., map the identity view on $[n]$ to the singleton
$1$. The only exceptions are variable applications, which are mapped
to the empty presheaf.

The translation extends to global processes as follows. Fixing a
global process $Q = (\recin{x_1 (\Del_1) \eq P_1, \ldots, x_k (\Del_k) \eq
  P_k}{P})$ typed in $\Gam$ with $n$ names, define the sequence
$(P^i)_{i \in \FinOrd_0}$ of open processes (all typed in $\Xi;\Gam$)
as follows.  First, $P^0 = P$. Then, let $P^{i+1} = \derivmap P^i$,
where $\derivmap$ is the \emph{derivation} endomap on open processes
typed in any extension $\Xi; (\Gam,\Del)$ of $\Xi;\Gam$, which unfolds
one layer of recursive definitions. This map is defined by induction
on its argument as follows:
 $$\begin{array}[t]{r@{\ =\ }lll}
\derivmap(x_l (a_1, \ldots, a_{k_l})) & P_l \subs{b_j \mapsto a_j}_{1 \leq j \leq k_l} \\
  \derivmap(P|Q) & {\derivmap P} | {\derivmap Q} 
  \end{array} \hspace*{.3cm}
  \begin{array}[t]{r@{\ =\ }lll}
    \derivmap(\nu a.P) &  \nu a. \derivmap P \\
    \derivmap(\sum_{i \in I} \alpha_i.P_i) &  \sum_{i \in I} \alpha_i.(\derivmap P_i),
\end{array}$$
where for all $l \in \ens{1, \ldots, k}$, $\Del_l = (b_1, \ldots,
b_{k_l})$, and $P\subs{\sigma}$ denotes simultaneous, capture-avoiding
substitution of names in $P$ by $\sigma$.

By construction, the translations of these open processes form a
sequence $\transl{P^0} \into \transl{P^1} \ldots$ of inclusions in
$\OPsh {\CV_n}$, such that for any natural number $i$ and view $V \in
\CV_n$ of length $i$, $\transl{P^j} (V)$
is fixed after $j = (k+1)i$, at worst, i.e., for all $j \geq (k+1)i$,
$\transl{P^j} (V) = \transl{P^{(k+1)i}} (V)$.  Thus, this sequence has a
colimit in $\OPsh {\CV_n}$, the presheaf sending any view $V$ of
length $i$ to $\transl{P^{(k+1)i}} (V)$. We put:
\begin{defn}\label{def:transl}
  Let the translation of $Q$ be $\transl{Q} = \colim_{i \in \FinOrd}
  \transl{P^i}$.
\end{defn}

Which equivalence is induced by this mapping on CCS, especially when
taking into account the interactive equivalences developed in the next
section? This is the main question we will try to address in future
work.

\section{Interactive equivalences}\label{sec:inter}
\subsection{Fair testing vs.\ must testing: the standard case}\label{ssec:inter:std}
An important part of concurrency theory consists in studying
\emph{behavioural equivalences}. Since each such equivalence is
supposed to define when two processes behave the same, it might seem
paradoxical to consider several of them.  Van
Glabbeek~\cite{DBLP:conf/concur/Glabbeek90} argues that each
behavioural equivalence corresponds to a physical scenario for
observing processes.

A distinction we wish to make here is between \emph{fair} scenarios,
and \emph{potentially unfair} ones. An example of a fair scenario is
when parallel composition of processes is thought of as modelling
different physical agents, e.g., in a game with several players.
Otherwise said, players are really independent. On the other hand, an
example of a potentially unfair scenario is when parallelism is
implemented via a scheduler.

This has consequences on so-called \emph{testing}
equivalences~\cite{DBLP:journals/tcs/NicolaH84}. Let $\tick$ be a
fixed action.
\begin{defn}
  A process $P$ is \emph{must orthogonal} to a context $C$, notation
  $P \mathrel\bot^m C$, when all maximal traces of $C[P]$ play $\tick$ at some
  point.
\end{defn}
Here, maximal means either infinite or finite without extensions.
Let $P^{\bot^m}$ be the set of all contexts must orthogonal to $P$.

\begin{defn}
  $P$ and $Q$ are \emph{must equivalent}, notation $P \sim_m Q$,
  when $P^{\bot^m} = Q^{\bot^m}$.
\end{defn}

In transition systems, or automata, we have $\Omega \sim_m
\Omega|\abar$ (where $\Omega$ is the looping process, producing
infinitely many silent transitions). This might be surprising, because
the context $C = a.\tick \para \trou$ intuitively should distinguish
these processes, by being orthogonal to $\Omega|\abar$ but not to
$\Omega$ alone. However, it is not orthogonal to $\Omega|\abar$,
because $C[\Omega|\abar]$ has an infinite looping trace giving
priority to $\Omega$.  This looping trace is unfair, because the
synchronisation on $a$ is never performed.  Thus, one may view the
equivalence $\Omega \sim_m \Omega|\abar$ as exploiting potential
unfairness of a hypothetical scheduler. 

Usually, concurrency theorists consider this too coarse, and resort to
\emph{fair} testing equivalence.
\begin{defn}
  A process $P$ is \emph{fair orthogonal} to a context $C$, notation
  $P \mathrel\bot^f C$, when all finite traces of $C[P]$ extend to
  traces that play $\tick$ at some point.
\end{defn}
Again, $P^{\bot^f}$ denotes the set of all contexts fair orthogonal to
$P$.
\begin{defn}
  $P$ and $Q$ are \emph{fair equivalent}, notation $P \sim_f Q$, when
  $P^{\bot^f} = Q^{\bot^f}$.
\end{defn}
This solves the issue, i.e., $\Omega \nsim_f \Omega|\abar$.

In summary, the mainstream setting for testing equivalences relies on
traces; and the notion of maximality for traces is intrinsically
unfair. This is usually rectified by resorting to fair testing
equivalence over must testing equivalence.  Our setting is more
flexible, in the sense that maximal plays are better behaved than
maximal traces. In terms of the previous section, this allows viewing
the looping trace $\Omega|\abar|a.\tick \xto{\tau} \Omega|\abar|a.\tick \xto{\tau}
\ldots$ as non-maximal. In the next sections, we define an abstract
notion of interactive equivalence (still in the particular case of CCS
but in our setting) and we instantiate it to define and study the
counterparts of must and fair testing equivalences.

\subsection{Interactive equivalences}
\begin{defn}
  A play is \emph{closed-world} when it is a composite of closed-world extended moves.
\end{defn}
Equivalently, a play is closed-world when all of its basic moves are
part of a closed-world move.

Let $\CW \into \CE$ be the full subcategory of closed-world plays,
$\CWofX$ being the \emph{fibre} over $X$ for the projection functor
$\CW \to \Eh$, i.e., the subcategory of $\CW$ consisting of
closed-world plays with base $X$, and morphisms $(\id_X, k)$ between
them\footnote{This is not exactly equivalent to what could be noted
  $\CW_X$, since in the latter there are objects $U \otni Y \into X$
  with a strict inclusion $Y \into X$. However, both should be
  equivalent for what we do in this paper, i.e., fair and must
  equivalences.}.

Let the category of \emph{closed-world behaviours} on $X$ be the
category $\GlX = \CWofXhat$ of presheaves on $\CWofX$.  We may now
put:
  \begin{defn}
  An \emph{observable criterion} consists for all positions $X$, of a
  replete subcategory $\bbot_X \into \GlX$.
\end{defn}
Recall that $\bbot_X$ being replete means that for all $F \in \bbot_X$
and isomorphism $f \colon F \to F'$ in $\GlX$, $F'$ and $f$ are in
$\bbot_X$.  

An observable criterion specifies the class of `successful',
closed-world behaviours. The two criteria considered below are two
ways of formalising the idea that a successful behaviour is one in
which all accepted closed-world plays are `successful', in the sense
that some player plays the tick move at some point.

We now define interactive equivalences.  Recall that $[F, G]$ denotes
the amalgamation of $F$ and $G$, and that right Kan extension along
$\op{i_Z}$ induces a functor $\Ran_{\op{i_Z}} \colon \widehat{\CV_Z}
\to \widehat{\CE_Z}$. Furthermore, precomposition with the canonical
inclusion $j_Z \colon \CW (Z) \into \CE_Z$ induces a functor $j_Z^*
\colon \widehat{\CE_Z} \to \widehat{\CW (Z)}$.  Composing the two, we
obtain a functor $\SStoGG \colon \SS_Z \to \Gl_Z$:
$$\SS_Z = \widehat{\CV_Z} \xto{\Ran_{\op{i_Z}}} \widehat{\CE_Z} \xto{j_Z^*} \widehat{\CW (Z)} = \Gl_Z.$$

\noindent \begin{minipage}[t]{0.7\linewidth}
\begin{defn}\label{defn:orth}%\ \\[-.8em] \noindent
  For any innocent strategy $F$ on $X$ and any pushout square $P$ of
  positions as on the right, with $I$ consisting only of channels, let
  $F^{\bbot_P}$ be the class of all innocent strategies $G$ on $Y$
  such that $\SStoGG([F, G]) \in \bbot_Z$.
\end{defn}
\end{minipage}
\begin{minipage}[t]{0.28\linewidth}
  \vspace*{-1.3em}
  \begin{equation}
    \Diag(.3,.8){%
      \pbk{X}{Z}{Y} %
    }{%
      |(I)| I \& |(Y)| Y \\
      |(X)| X \& |(Z)| Z %
    }{%
      (I) edge (Y) %
      edge (X) %
      (X) edge (Z) %
      (Y) edge (Z) %
    }\label{eq:orth}
  \end{equation}
\end{minipage}
\vspace*{1.3em}

Here, $G$ is thought of as a \emph{test} for $F$. Also, $P$ denotes
the whole pushout square and $F^{\bbot_P}$ denotes all the valid tests
for the considered pushout square $P$.  From the CCS point of view,
$I$ corresponds to the set of names shared by the process under
observation $(F)$ and the test $(G)$.

\begin{defn} Any two innocent strategies $F,F'\in\SSX$ are
  \emph{$\bbot$-equivalent}, notation $F \sim_{\bbot} F'$, iff for all
  pushouts $P$ as in~\ref{eq:orth}, $F^{\bbot_P} = {F'}^{\bbot_P}$.
\end{defn}

\subsection{Fair vs.\ must}\label{subsec:fairmust}

Let us now define fair and must testing equivalences.  Let a
closed-world play be \emph{successful} when it contains a $\tickn$.
Furthermore, for any closed-world behaviour $G \in \GlX$ and
closed-world play $U \in \CWofX$, an \emph{extension} of a state
$\state \in G (U)$ to $U'$ is a $\statei \in G (U')$ with $i \colon U
\to U'$ and $G (i)(\statei) = \state$. The extension $\statei$ is
\emph{successful} when $U'$ is. The intuition is that the behaviour
$G$, before reaching $U'$ with state $\sigma'$, passed through $U$
with state $\sigma$.
\begin{defn}
  The \emph{fair} criterion $\bbot^f$ contains all closed-world
  behaviours $G$ such that any state $\state \in G (U)$ for finite $U$
  admits a successful extension.
\end{defn}

Now call an extension of $\state \in G (U)$ \emph{strict} when $U \to
U'$ is not surjective, or, equivalently, when $U'$ contains more moves
than $U$.  For any closed-world behaviour $G \in \GlX$, a state
$\state \in G (U)$ is $G$-\emph{maximal} when it has no strict
extension.
\begin{defn}
  Let the \emph{must} criterion $\bbot^m$ consist of all closed-world
  behaviours $G$ such that for all closed-world $U$ and $G$-maximal
  $\state \in G (U)$, $U$ is successful.
\end{defn}

As explained in the introduction and Section~\ref{ssec:inter:std},
unlike in the standard setting, this definition of must testing
equivalence distinguishes between the processes $\Omega$ and
$\Omega|\abar$. Indeed, take the CCS context $C = a.\tick \para
\trou$, which we can implement by choosing as a test the strategy $T =
\transl{a.\tick}$ on a single player knowing one channel $a$. Taking
$I$ to consist of the sole channel $a$, the pushout $Z$ as in
Definition~\ref{defn:orth} consists of two players, say $x$ for the
observed strategy and $y$ for the test strategy, sharing the channel
$a$.  Now, assuming that $\Omega$ loops deterministically, the global
behaviour $G = \SStoGG([\transl{P}, T])$ has exactly one state on the
identity play, and again exactly one state on the play $\paraof{1}$
consisting of only one fork move by $x$. Thus, $G$ reaches a position
with three players, say $x_1$ playing $\Omega$, $x_2$ playing $\abar$,
and $y$ playing $a.\tick$.  The play with infinitely many silent moves
by $x_1$ is not maximal: we could insert (anywhere in the sequence of
moves by $x_1$) a synchronisation move by $x_2$ and $y$, and then a
tick move by the avatar of $y$. Essentially: our notion of play is
more fair than just traces.

To get more intuition about must testing equivalence in our setting,
we prove that it actually coincides with the testing equivalence
generated by the following criterion:
\begin{defn}
  The \emph{spatially fair} criterion $\bbot^{sf}$ contains all
  closed-world behaviours $G$ such that any state $\state \in G (U)$
  admits a successful extension.
\end{defn}
This criterion is almost like the fair criterion, except that we do
not restrict to finite plays. The key result to show the equivalence is:
\begin{thm}\label{thm:spatialfairmust}
  For any innocent strategy $F$ on $X$, any state $\state \in
  \SStoGG(F) (U)$ admits a $\SStoGG(F)$-maximal extension.
\end{thm}
The proof is in Appendix~\ref{sec:maxext}.  Thanks to the theorem, we
have:
\begin{lem}\label{lem:spatialfairmust}
  For all $F \in \SS_X$, $\SStoGG(F) \in \bbot^m_X$ ~iff~ $\SStoGG(F)
  \in \bbot^{sf}_X$.
\end{lem}
\begin{myproof}
  Let $G = \SStoGG (F)$.

  \emph{($\Rightarrow$)} By Theorem~\ref{thm:spatialfairmust}, any
  state $\state \in G (U)$ has a $G$-maximal extension $\statei \in G
  (U')$, which is successful by hypothesis, hence $\state$ has a
  successful extension.

  \emph{($\Leftarrow$)} Any $G$-maximal $\state \in G (U)$ admits by
  hypothesis a successful extension which may only be on $U$ by
  $G$-maximality, and hence $U$ is successful.
\end{myproof}

(Note that $U$ is not necessarily finite in the proof of the right-to-left
implication, so that the argument does not apply to the fair criterion.)

Now comes the expected result:
\begin{thm}
For all $F, F' \in \SS_X$, $F \sim_{\bbot^m} F'$ iff $F \sim_{\bbot^{sf}} F'$.
\end{thm}
\begin{myproof}  
  \emph{($\Rightarrow$)} Consider two innocent strategies $F$ and $F'$
  on $X$, and an innocent strategy $G$ on $Y$ (as in the
  pushout~\eqref{eq:orth}). As in spatial decomposition
  (Proposition~\ref{prop:spatial}), copairing induces an isomorphism
  $\SS_X \times \SS_Y \to \SS_Z$, and we have, using
  Lemma~\ref{lem:spatialfairmust}:
  \begin{align*}
    \SStoGG[F, G] \in \bbot^{sf} & \quad\text{iff}\quad
    \SStoGG[F, G] \in \bbot^m \\ &\quad\text{iff}\quad 
    \SStoGG[F', G] \in \bbot^m \\ &\quad\text{iff}\quad
    \SStoGG[F', G] \in \bbot^{sf} 
  \end{align*}
%   \begin{center}
% %    \begin{tabular}{ll}
% %    $\SStoGG(F \glue T) \in \bbot^f$ & iff $\SStoGG(S \glue T) \in \bbot^m$ \\ & iff $\SStoGG(S'
% %    \glue T) \in \bbot^m$ \\ & iff $\SStoGG(S' \glue T) \in \bbot^f$.
% %  \end{tabular}
%     $\SStoGG(F \glue G) \in \bbot^{sf}$ \hfil iff \hfil $\SStoGG(F
%     \glue G) \in \bbot^m$ \hfil iff \hfil $\SStoGG(F' \glue G) \in
%     \bbot^m$ \hfil iff \hfil $\SStoGG(F' \glue G) \in \bbot^{sf}$.
%   \end{center}
  \emph{($\Leftarrow$)}  Symmetric.
\end{myproof}

Intuitively, must testing only considers spatially fair schedulings, in
the sense that all players appearing in a play should be given the
opportunity to play: no one should starve.

However, this is not the only source of unfairness, so that must
testing and fair testing differ. To see this, consider the CCS process
$P=\nu b.\recin{x (a,b)\eq \bbar | (b.(x (a,b)) +\abar)}{x (a,b)}$,
that can repeatedly perform synchronisations on the private channel
$b$, until it chooses to perform an output on $a$.  We have
$\transl{\Omega}\sim^{sf}\transl{P}$ while
$\transl{\Omega}\not\sim^{f}\transl{P}$. Indeed, since the choice
between doing a synchronisation on $b$ or an output on $a$ is done by
a single player, the infinite play where the output on $a$ is never
performed is maximal: no player starve, we just have a player that
repeatedly chooses the same branch, in an unfair way.

We leave for future work the investigation of such unfair scenarios
and their correlation to the corresponding behaviours in classical
presentations of CCS.

%\bigskip
% \paragraph{Acknowledgements}
% \dam{j'aurais tendance à virer les remerciements pour la soummission, non?}
% \tom{allez.}
% Thanks to the courageous having endured
% the first versions of this work. Special thanks to Paul-Andr\'e
% Melli\`es for his graphical design skills, and to pseudonymous referee
% Michel Houellebecq of ICE '11, not only for our very useful and
% enjoyable discussion, but also for tolerating our rather poor
% literary style. Finally, thanks to Richard Garner and Mark Weber for
% categorical advice.
% 
% 
% 

\appendix
\section{Temporal decomposition}\label{sec:proofEE}
This section is a proof of Theorem~\ref{thm:temp}.  Let us first
review the general equivalences mentioned in the proof sketch.  The
product of a family of presheaf categories is isomorphic to the
category of presheaves over the corresponding coproduct of categories:
\begin{lem}
  We have $\prod_{M \in \MMM_n} \SS_{\cod (M)} \iso [\op{\sum_{M \in \MMM_n} \CV_{\cod (M)}},\Set]$.
\end{lem}

Furthermore, let the functor $\Delta \colon \Set \to \Chat$ map any
set $X$ to the constant presheaf mapping any $C \in \C$ to $X$.  We
have:
\begin{lem}
  For any small category $\C$, $\Fam (\Chat) \equi (\Chat \downarrow \Delta)$.
\end{lem}
\begin{myproof}
  A generalisation of the more well-known $\Set^X \equi \Set / X$.
\end{myproof}
\begin{cor} We have:
  $$\Fam \left ( \prod_{M \in \MMM_n} \SS_{\cod (M)} \right ) \equi
  ([\op{\sum_{M \in \MMM_n} \CV_{\cod (M)}},\Set] \downarrow
  \Delta).$$
\end{cor}

We now construct the lax pushout~\eqref{eq:laxpsht}.  A first step is
the construction, for each move $[n] \into M \otni [n']$, of a functor
$(- \rond M) \colon \CVni \to \CVn$ given by precomposition with $M$
in $\Cospan{\Chat}$.  This functor maps any $V_1 \colon [n'] \into
V_1$ to the view $V_1 \rond M$, i.e., the view $[n] \into V'_1$
defined by the colimit
\begin{center}
  \Diag{%
    \pbk{V}{V''}{V'} %
  }{%
    |(n)| [n] \& \& |(n')| [n'] \\
    \& |(V)| M \& \& |(V')| V_1 \\
    \& \& |(V'')| V'_1. %
  }{%
    (n) edge[into] (V) %
    (n') edge[linto] (V) %
    edge[into] (V') %
    (V) edge (V'') %
    (V') edge (V'') %
  }
\end{center}
This of course relies on the choice of such a colimit for every $V$
and $V_1$.  Any morphism $f \colon V_1 \to V_2$ in $\CV_{[n']}$,
letting $V'_2 = V_2 \rond V$, is mapped to the dashed morphism induced by
universal property of pushout in
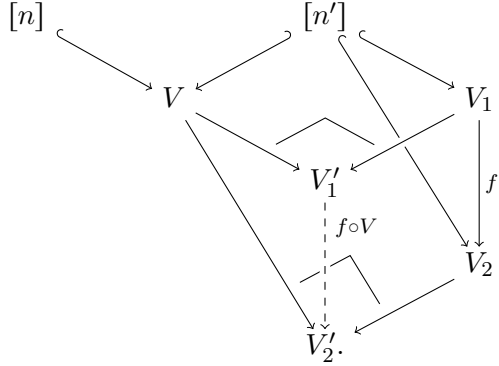
\begin{center}
  \Diag{%
    \pbk{V}{V'_1}{V_1} %
    \pbk{V}{V'_2}{V_2} %
    \path[->,draw] (V'_1) edge[dashed,fore] node[pos=.2,anchor=west] {$\scriptstyle f \rond V$} (V'_2) %
    ; %
  }{%
    |(n)| [n] \& \& |(n')| [n'] \\
    \& |(V)| V \& \& |(V_1)| V_1 \\
    \& \& |(V'_1)| V'_1 \& \\
    \& \& \& |(V_2)| V_2 \\
    \& \& |(V'_2)| V'_2. \&  %
  }{%
    (n) edge[into] (V) %
    (n') edge[linto] (V) %
    edge[into] (V_2) %
    edge[into] (V_1) %
    (V) edge (V'_1) %
    (V_1) edge[fore] (V'_1) %
    edge[labelr={f}] (V_2) %
    (V) edge (V'_2) (V_2) edge (V'_2) %
  }
\end{center}
Once the choice has been made on objects, the map for morphisms is determined
uniquely. 

This family of functors allows us to decompose $\CVn$ as follows:
\begin{lem}
The diagram
  \begin{equation}
    \Diag (1,1) {%
      \twocell[.5]{Vnsi}{Vn}{un}{}{cell=0,bend
        right,labelal={\lambda}} %
    }{%
      |(Vns)| {\sum_{M \in \MMM_n} \op\CV_{\cod (M)}} \&  |(Vnsi)| {\sum_{M \in \MMM_n} \op\CV_{\cod (M)}} \\
      |(un)| 1 \& |(Vn)| \op\CVn %
    }{%
      (Vns) edge[identity] (Vnsi) %
      edge[shorten <=.1cm,labell={!}] (un) %
      (un) edge[labeld={\name{\id_{[n]}}}] (Vn) %
      (Vnsi) edge[shorten <=.1cm,labelr={[- \rond M]_{M \in
          \MMM_n}}] (Vn) %
    }\label{eq:lpo}
  \end{equation}
  is a lax pushout, where $\lambda_{M, V} \colon \id_{[n]} \to M \rond
  V$, seen in $\CVn$, is the obvious inclusion.
\end{lem}
\begin{myproof}
  For any category $\C$, taking such a lax pushout of $\id_\C$ with
  $1$ just adds a terminal object to $\C$. The rest is an easy
  verification. A dual result of course holds with $\CVn$, reversing
  the direction of $\lambda$.
\end{myproof}

Now, it is well-known that, in any small 2-category $\K$, any
contravariant hom-2-functor, i.e., 2-functor of the shape $\K (-, X)$
for $X \in K$, maps weighted colimits in $\K$ to weighted limits in
$\Cat$. For an introduction to weighted limits and colimits in the
case of enrichment over $\Cat$, see Kelly~\cite{Kelly89}. Here, for
any 2-category $P$, and 2-functors $G \colon P \to \K$ and $J \colon
\op P \to \Cat$, any colimit $L = J \star G$ of $G$ weighted by $J$
with unit $\xi \colon J \to \K (G (-), L)$ in $[\op P, \Cat]$ is
mapped, for any object $X \in \K$, by the hom-2-functor $\K (-, X)$ to
a limit of $\K (G (-), X) \colon \op P \to \Cat$ weighted by $J$ in
$\Cat$, with unit $\K (\xi,X) \colon J \to \Cat (\K (L,X), \K (G (-),
X))$, in $\Cat$. In particular, lax pushouts are mapped to lax
pullbacks. As usual, considering a larger universe, we may replace
$\Cat$ with $\CAT$ and obtain the same results with $\K = \Cat$.  

Recalling our lax pushout~\eqref{eq:lpo} and taking the hom-categories
to $\Set$, we obtain a lax pullback
  \begin{center}
    \Diag (1,1) {%
      \path[<-,draw] %
      (Vns) edge[identity] (Vnsi) %
      edge[shorten <=.1cm,labell={!^*}] (un) %
      (un) edge[labeld={}] (Vn) %
      (Vnsi) edge[shorten <=.1cm,labelr={}] (Vn) %
      ; %
      \twocell[.5]{Vnsi}{Vn}{un}{}{cell=0,bend
        right,labelal={\lambda^*}} %
    }{%
      |(Vns)| [\op{\sum_{M \in \MMM_n} \CV_{\cod (M)}},\Set] \&  |(Vnsi)|  [\op{\sum_{M \in \MMM_n} \CV_{\cod (M)}},\Set] \\
      |(un)| \Set \& |(Vn)| \SSn %
    }{%
    }
  \end{center}
  in $\CAT$, i.e., a comma category. But observe that restriction
  along $!$ is precisely $\Delta \colon \Set \to [\op{\sum_{M \in
      \MMM_n} \CV_{\cod (M)}},\Set]$, so we have indeed shown that $\SSn$
  is a comma category $[\op{\sum_{M \in \MMM_n} \CV_{\cod (M)}},\Set]
  \downarrow \Delta$.

\section{Maximal extensions}\label{sec:maxext}
This section is a proof of Theorem~\ref{thm:spatialfairmust}.  

\begin{lem}
  For any position $X$, the category $\CWofX$ of closed-world plays is a
  preorder.
\end{lem}
\begin{myproof}
  Easy.
\end{myproof}

In the following, we consider the quotient poset.
\begin{lem}
  In $\CWofX$, any non-decreasing chain admits an upper bound.
\end{lem}

Recall $\MMM$, the graph of all basic moves, and the set $\MMM_n$ of
edges from $n$, for each $n$.  Let now, for each $n$, $\MMM^f_n$ be
the analogous set with full moves, i.e., the set of isomorphism
classes of full moves from $[n]$.
\begin{lem}\label{lem:jointsmaps}
  For each play $U \in \CEX$, the coproduct of all $s$ maps from full moves
  \begin{equation}
  \left ( \sum_{n \in \FinOrd} \sum_{M \in \MMM^f_n} U (M) \right ) \to \sum_{n \in \FinOrd} U[n],\label{eq:smaps}
\end{equation}
  is injective.
\end{lem}
Recall here that for forking, we have also called $s$ the common
composite $l \rond s = r \rond s$ (see the discussion following
Definition~\ref{def:C3}).
\begin{myproof}
  By induction on $U$.
\end{myproof}

\begin{lem}\label{lem:pourzorn}
  Any non-decreasing sequence in the poset $\CWofX$ admits its
  colimit in $\Chat$ as an upper bound.
\end{lem}
\begin{myproof}
  Consider any increasing sequence $U^1 \into U^2 \into \ldots$ of
  plays in $\CWofX$. Let $U$ be its colimit in $\Chat$. We want to
  prove that $U$ is a play.

  First, observe that $U$ satisfies joint injectivity of $s$-maps as
  in Lemma~\ref{lem:jointsmaps}: indeed, if we had a player $p$ and
  two full moves $M$ and $M'$ such that $s (M) = s (M') = p$, then all
  of $M$, $M'$, and $p$ would appear in some $U^i$, which, being a
  play, has to satisfy joint injectivity.

  For each $n$, $U^n$ comes with a sequence of compatible
  (closed-world) extended moves
  $$X = X^n_0 \into M^n_1 \otni X^n_1 \into \ldots \otni X^n_{i-1} \into M^n_{i} \otni X^n_i \into \ldots$$
  which are also (by the colimit cocone) morphisms over $U$ in
  $\Chat$. For each $i \geq 1$, taking the colimit of the $i$ first
  moves yields a finite play $X \into U^n_i \otni X^n_i$. By
  convention, letting $U^n_0 = X$ extends this to $i \geq
  0$. Similarly, we may consider all the given plays infinite, by
  accepting not only extended moves, but also identity cospans.

  We consider the poset of pairs $(N, n) \in \ens{(0,0)} \uplus
  \sum_{N \in \FinOrd^*} N$, with lexicographic order, i.e., $(N, n)
  \leq (N',n')$ when $N < N'$ or when $N = N'$ and $n \leq n'$.

  We will construct by induction on $(N,n)$ a sequence of composable
  closed-world moves, with colimit $U'$, such that for all $(N,n)$,
  $U^n_{N-n+1} \subseteq U'$ in $\CWofX / U$. More precisely, we
  construct for each $(N,n)$ an integer $K_{N,n}$ and a sequence
  $$X = X^{N,n}_0 \into M^{N,n}_1 \otni X^{N,n}_1 \into  \ldots 
  \otni X^{N,n}_{K_{N,n} - 1} \into M^{N,n}_{K_{N,n}} \otni
  X^{N,n}_{K_{N,n}},$$ (again, if $K_{N,n} = 0$, we mean the empty
  sequence) such that
  \begin{itemize}
  \item for all $(N',n') < (N,n)$, we have $K_{N',n'} \leq K_{N,n}$
    and the sequence $(M^{N',n'}_i)_{i \in K_{N',n'}}$ is a prefix of
    $(M^{N,n}_{i \in K_{N,n}})$;
  \item and the colimit, say $U_{N,n}$, of $(M^{N,n}_i)_{i \in
      K_{N,n}}$ is such that for all $(N',n') \leq (N,n)$, $U^{n'}_{N-n'+1}
    \subseteq U_{N,n}$ in $\CWofX / U$.
  \end{itemize}
  For the base case, we let $K_{0,0} = 0$, which forces $M^{0,0}$ to
  be the empty sequence on $X$.

  For the induction step, consider any $(N,n) \neq (0,0)$, and let
  $(N_0,n_0)$ be the predecessor of $(N,n)$. The induction hypothesis
  gives a $K_{N_0,n_0}$ and a sequence $(M^{N_0,n_0}_i)_{i \in
    K_{N_0,n_0}}$ satisfying some hypotheses, among which the
  existence of a diagram
  \begin{center}
    \diag{%
      |(X)| X \& |(UnNn)| U^n_{N-n} \& |(XnNn)| X^n_{N-n} \& |(MnNni)| M^n_{N-n+1} \& |(XnNni)|  X^n_{N-n+1} \\
      |(Xi)| X \& |(UN0n0)| U_{N_0,n_0} \& |(XN0n0)| X^{N_0,n_0}_{K_{N_0,n_0}} %
    }{%
      (X) edge (UnNn) %
      edge[identity] (Xi) %
      (Xi) edge (UN0n0) %
      (XnNn) edge (UnNn) %
      edge (MnNni) %
      (XnNni) edge (MnNni) %
      (XN0n0) edge (UN0n0) %
      (UnNn) edge (UN0n0) %
    }
  \end{center}
  over $U$.  

  Now, if $M^n_{N-n+1} \to U$ factors through $U_{N_0,n_0}$, then we
  put $K_{N,n} = K_{N_0,n_0}$ and $(M^{N,n}_i)_{i \in K_{N,n}} =
  (M^{N_0,n_0}_i)_{i \in K_{N_0,n_0}}$, and all induction hypotheses
  go through.

  Otherwise, $M^n_{N-n+1}$ is played by players in $X^n_{N-n}$ which
  are not in the joint image of all $s$ maps~\eqref{eq:smaps} in
  $U_{N_0,n_0}$, otherwise $s$ maps in $U$ could not be jointly
  injective, contradicting Lemma~\ref{lem:jointsmaps}. Technically, the diagram
  $$X^n_{N-n} \to M^n_{N-n+1} \ot X^n_{N-n+1}$$
  is obtained by pushing some (non-extended) closed-world move $Y \to
  M \ot Y'$ along some morphism $I \to Z$ from an interface $I$, and
  the induced morphism $Y \to X^n_{N-n} \to U^n_{N-n} \to U_{N_0,n_0}$
  factors through $X^{N_0,n_0}_{K_{N_0,n_0}}$. We consider the
  subposition $Z' \subseteq X^{N_0,n_0}_{K_{N_0,n_0}}$ making
  \begin{center}
    \diag{%
      |(I)| I \& |(Y)| Y \\
      |(Z')| Z' \& |(X)| X^{N_0,n_0}_{K_{N_0,n_0}} %
    }{%
      (I) edge[into] (Y) edge (Z') %
      (Z') edge[into] (X) %
      (Y) edge (X) %
    }
  \end{center}
  a pushout; $Z'$ consists of the players in
  $X^{N_0,n_0}_{K_{N_0,n_0}}$ that are not in the image of $Y$, plus
  their names, plus possibly missing names from $I$. 

  Then, pushing $Y \to M \ot Y'$ along $I \to Z'$, we obtain an
  extended move $X^{N_0,n_0}_{K_{N_0,n_0}} \into M' \otni X'$. We let
  $K_{N,n} = K_{N_0,n_0} + 1$ and define $(M^{N,n}_i)_{i \in K_{N,n}}$
  to be the extension of $(M^{N_0,n_0}_i)_{i \in K_{N_0,n_0}}$ by
  $M'$. This induces a unique map $U_{N,n} \to U$ by universal
  property of $U_{N,n}$ as a colimit. All induction hypotheses go
  through; in particular, $U^n_{N-n+1}$ is a union $U^n_{N-n} \cup
  M^n_{N-n+1}$ in $\CWofX / U$, and actually a union $U^n_{N-n} \cup
  M$; similarly, $U_{N,n} = U_{N_0,n_0} \cup M$; so, since we have
  $U^n_{N-n} \subseteq U_{N_0,n_0}$ by induction hypothesis, we obtain
  $U^n_{N-n+1} \subseteq U_{N,n}$.

  The sequences $M^{N,n}$ induce by union a possibly infinite sequence
  of closed-world extended moves, i.e., a closed-world play $U'$, such
  that for all $(N,n)$, $U^n_{N-n+1} \subseteq U'$, hence, for all
  $n$, $U^n \subseteq U' \subseteq U$, i.e., $U' \iso U$. Thus, $U$ is
  indeed a play.
\end{myproof}

We are almost ready for proving Theorem~\ref{thm:spatialfairmust}. We
just need one more lemma.  Consider any innocent strategy $F$ on $X$,
play $U \in \CW (X)$, and any state $\state \in \SStoGG(F) (U)$.
Consider now the poset $F_\state$ of $\SStoGG (F)$-extensions of
$\state$ (made into a poset by choosing a skeleton of $\CWofX$), where
$\state' \in F (U') \leq \state'' \in F (U'')$ iff $U' \leq U''$. This
poset is not empty, since it contains $\state$. Furthermore, we have:
\begin{lem}
  Any non-decreasing sequence in $F_\state$ admits an upper bound.
\end{lem}
\begin{myproof}
  Any such sequence, say $(\state_i)_{i \in \FinOrd}$, induces a
  non-decreasing sequence of plays in $\CWofX$, say $(U_i)_{i}$, which
  by Lemma~\ref{lem:pourzorn} admits its colimit, say $U'$, as an
  upper bound. Now, any view inclusion $j \colon V \into U'$, factors
  through some $U_i$, and we let $\state_{j} = \restr{(\state_i)}{V}$
  (this does not depend on the choice of $i$). This assignment
  determines (by innocence of $F$ and by construction of the right Kan
  extension as an end) an element $\state' \in F (U')$, which is an
  upper bound for $(\state_i)_{i \in \FinOrd}$.
\end{myproof}

\begin{myproof}[Proof of Theorem~\ref{thm:spatialfairmust}]
  Consider any innocent strategy $F$ on $X$, play $U \in \CW (X)$, and
  any state $\state \in \SStoGG(F) (U)$.  Consider as above the poset
  $F_\state$ of $\SStoGG (F)$-extensions of $\state$. By the last
  lemma, we may apply Zorn's lemma to choose a maximal element of
  $F_\state$, which is a $\SStoGG(F)$-maximal extension of $\state$.
\end{myproof}

% if you use BibTeX then
 \bibliography{common/bib}
% else

\hypertarget{lastpage}{}
\end{document}